\documentclass[12pt]{article}
\usepackage{graphicx}
\usepackage{subfigure}
\usepackage{float}
\usepackage{amsmath}
\usepackage{bm}
\usepackage{times}
\usepackage{color,soul}
\usepackage[utf8x]{inputenc}
\usepackage[T1]{fontenc}
\usepackage[euler]{textgreek}
\usepackage[right]{lineno}
\PassOptionsToPackage{hyphens}{url}\usepackage{hyperref}
\usepackage[dvipsnames]{xcolor}
\usepackage{floatpag}
\floatpagestyle{plain}
\usepackage{abstract}

\usepackage[super,comma]{natbib}
\newcommand{\norm}[1]{\left\lVert#1\right\rVert}

\title{\vspace{-1in}Fooling the primate brain with\\ minimal, targeted image manipulation}

\author{
    \hspace{-0.25in}
    Li Yuan,\textsuperscript{1,2,7} Will Xiao,\textsuperscript{1,3,7,*} Giorgia Dellaferrera,\textsuperscript{4} Gabriel Kreiman,\textsuperscript{4,5}
    \\ \hspace{-0.25in}
    Francis E.H. Tay,\textsuperscript{6} Jiashi Feng,\textsuperscript{2} Margaret S. Livingstone\textsuperscript{1,*} 
    \\ \hspace{-0.25in}
    \small{\textsuperscript{1}Department of Neurobiology, Harvard Medical School, Boston, MA 02115, U.S.A.} \\ \hspace{-0.25in}
    \small{\textsuperscript{2}Department of Electrical and Computer Engineering, National University of Singapore, Singapore 117583} \\ \hspace{-0.25in}
    \small{\textsuperscript{3}Department of Molecular and Cellular Biology, Harvard University, Cambridge, MA 02134, U.S.A.} \\ \hspace{-0.25in}
    \small{\textsuperscript{4}Boston Children's Hospital, Harvard Medical School, Boston, MA 02115, U.S.A.} \\ \hspace{-0.25in}
    \small{\textsuperscript{5}Center for Brains, Minds and Machines, Cambridge, MA 02115, U.S.A.} \\ \hspace{-0.25in}
    \small{\textsuperscript{6}Department of Mechanical Engineering, National University of Singapore, Singapore 117583} \\ \hspace{-0.25in}
    \small{\textsuperscript{7}These authors contributed equally} \\ \hspace{-0.25in}
    \small{\textsuperscript{*}To whom correspondence should be addressed;} \\ \hspace{-0.25in}
    \small{E-mail: xiaow@fas.harvard.edu; margaret\_livingstone@hms.harvard.edu}
}

\begin{document}
\date{}
\maketitle
\linenumbers
\pagestyle{plain}
\nocite{*}

\section*{Abstract}
Artificial neural networks (ANNs) are considered the current best models of biological vision. ANNs are the best predictors of neural activity in the ventral stream; moreover, recent work has demonstrated that ANN models fitted to neuronal activity can guide the synthesis of images that drive pre-specified response patterns in small neuronal populations. Despite the success in predicting and steering firing activity, these results have not been connected with perceptual or behavioral changes. Here we propose an array of methods for creating minimal, targeted image perturbations that lead to changes in both neuronal activity and perception as reflected in behavior. We generated `deceptive images' of human faces, monkey faces, and noise patterns so that they are perceived as a different, pre-specified target category, and measured both monkey neuronal responses and human behavior to these images.
We found several effective methods for changing primate visual categorization that required much smaller image change compared to untargeted noise.
Our work shares the same goal with adversarial attack, namely the manipulation of images with minimal, targeted noise that leads ANN models to misclassify the images.
Our results represent a valuable step in quantifying and characterizing the differences in perturbation robustness of biological and artificial vision.

\clearpage

\baselineskip24pt

\section*{Introduction}
Deep artificial neural networks (ANNs), which mimic the approximately hierarchical structure of the visual cortex \citep{Kreiman21_vision}, currently constitute the most accurate models of biological vision \citep{yamins2014performance, cadena2019deep,schrimpf2018brain}. ANNs not only are good predictors of ventral stream neural activity, but also can be used to design images that drive prescribed patterns of neuronal responses, thus achieving a kind of neural activity control \citep{bashivan2019neural,walker2019inception}. Bashivan et al. \cite{bashivan2019neural} used an AlexNet\cite{alexnet}-based encoding model to synthesize images to test control in two settings: In the `neural stretch' setting, the model was used to design synthetic images that activated specific neurons more strongly than hundreds of naturalistic stimuli. In `neural population state control,' synthetic images highly activated one neuron while avoided activating other simultaneously recorded neurons, achieving such `one-hot' sparse response patterns again more closely than hundreds of naturalistic images.
In related work, Walker et al. \cite{walker2019inception} used an ANN predictive model of mouse V1 responses to synthesize `most exciting inputs,' which frequently drove neuronal responses more strongly than optimized Gabor filters.

Thus, prior work demonstrates the exceptional explanatory power of ANN-based models of biological vision. However, while prior results achieved remarkable control of the activity of small neuron populations, these results do not explore the perceptual consequences of such control, nor aim to control neuronal responses toward driving a specific behavior. Here, we built upon prior work and designed several approaches to image synthesis---including methods based on an ANN encoding model of neuronal activity---to steer both neuronal responses and behavior in a binary image categorization task. We generated `deceptive images' that, (i) when presented to macaques, are miscategorized by face-selective neurons as the target category; and that, (ii) when presented to human observers, mislead their perception during a visual categorization task.  We focused on the well-characterized face-processing network \citep{moeller2008patches} in macaque inferior temporal cortex (IT). Correspondingly, we investigated three directions of image manipulation by creating deceptive images of human faces to look like monkey faces, deceptive images of monkey faces to look like human faces, and deceptive images of noise patterns to look like human faces.

We designed and tested several methods for generating deceptive images, including linear interpolation toward the target class (without or with affine alignment); CycleGAN translation \citep{zhu2017unpaired}; generating adversarial images for an ensemble of ANNs \citep{elsayed2018adversarial}; and `gray box' methods using an ANN-based predictive model of IT neuronal responses (thus, utilizing partial knowledge about primate vision). We produced deceptive images that specifically changed primate categorization at very low noise levels, when non-specific noise\citep{geirhos2018generalisation} barely affected categorization. In the most difficult misclassification direction---human-to-monkey---the gray box and CycleGAN-based methods were the most effective. In the other two directions, straightforward linear interpolation toward the target class was the most effective.


Our deceptive images are designed to be minimally modified (within a limited budget for pixel value change) from one category to look like a target category. The notion of minimally perturbing an image to change its categorization is no new concept: convolutional neural networks (CNNs) have been found susceptible to similar adversarial attacks, the phenomenon that adding carefully crafted, minute noise to an image can cause ANNs to misclassify it with high confidence \citep{szegedy2013intriguing,goodfellow2014explaining}.  
Our characterization of deceptive images for primate vision supplies a much needed point of comparison between the robustness of biological and artificial vision. Across the wide variety of methods we tested, the noise threshold for changing categorization was much higher for primate vision than for ANNs. In particular, adversarial images designed for ANNs were highly effective for ANNs, as expected, but were essentially ineffective for humans and monkey neurons, even at the highest, easily perceivable noise level we tested. Finer-grained response patterns to deceptive images also revealed differences between primate and ANN vision. Monkey neurons and human behavior were similar, as correlated as possible within the ceiling imposed by individual variability. In contrast, model classification of deceptive images only weakly correlated with primate classification. Finally, ANN-based encoding models of neuronal responses, when fitted on clean image data, did not generalize to predict responses to deceptive images. Our characterization of the difference between primate and computer vision in response to category-specific noise provides useful clues for building better models of the brain and more robust computer vision algorithms. Furthermore, the image synthesis methods we developed to flip the perceived image category opens up a new experimental paradigm for studying neuronal processing and its link to behavior.

\section*{Results}

\subsection*{Deceptive Images of Human Faces Evoked Monkey Face-like Neuronal Responses}

We started by designing deceptive images of human face images that would elicit monkey face-like responses (\emph{human→monkey} or \emph{h2m} manipulation) in face-selective monkey neurons. Because the minimal image changes that change image category for primate vision are virtually unexplored, we tested a wide range of image manipulation noise strengths (\emph{noise levels}) and a variety of perturbation methods.

To choose a noise range that was likely to progress from unsuccessful to successful semantic flip, we reasoned that image category is trivially changed if an original-class image is replaced by an image from the opposite class. Thus, we computed the distance distribution between unmodified human and monkey faces, considering 250 images in each category. Image distance was quantified by mean-squared error (MSE) using vectorized pixel values (range: 0--255). At constant image resolution, MSE is equivalent to the square of the $l_2$-norm (i.e., Euclidean distance), a common metric in the adversarial attack literature. We elected to use MSE instead of the $l_2$-norm because the latter depends on the number of pixels. The typical human-to-monkey pairwise distance in our image set was MSE = 6500 ± 2300 (mean ± stdev for all numbers in text unless otherwise noted; \textbf{Figure~\ref{fig:neural_sup1}b}). However, for successful semantic change, it suffices to replace a human face with the closest monkey face. With 250 images in each category, the minimum human-to-monkey distance was 2800 ± 700. We thus reasoned that an upper range of MSE = 800 was reasonable to test, as it was safely lower than the budget for simply replacing the image. We selected a lower range of MSE = 200 because this amount of image change was barely perceptible. We tested 10 evenly spaced noise levels covering this range of MSE.

The first manipulation method we considered, as a reference, was to linearly interpolate between a human face and the closest monkey face. This method is guaranteed to achieve complete success at a high enough noise level. Second, we included a slight modification by interpolating toward the closest monkey face up to an optimized affine transform. Other transformations, such as reducing contrast, can further reduce the distance to the target image. However, we did not test further transformations because an excessively transformed image will eventually cease to represent the target category.

Third, we included a method based on CycleGAN \cite{zhu2017unpaired}. CycleGAN can learn to translate between two image categories without any paired examples. CycleGAN does not explicitly optimize pixel-level proximity between the original and translated image. Nevertheless, the translated image is closely related to the original image, because the original (as opposed to another instance from the original class) can be approximately recovered from the translated image using a reverse translation. We trained a CycleGAN on several hundred human and monkey faces. Using the trained network, we translated each original human face into the monkey class, then linearly interpolated between the two at defined noise levels.

Fourth, we tested adversarial images created for an ensemble of convolutional neural networks (CNNs). Prior work \cite{elsayed2018adversarial} shows that adversarial images designed for CNNs can bias human perception, but only under severe viewing time limits and backward masking where accuracy on unmodified images is reduced to around 65\% (when chance is 50\%). We reproduced this method in our setting without the same viewing time limit. To create deceptive images between human and monkey faces, we used CNNs pre-trained on ImageNet \citep{deng2009imagenet} and fine-tuned them on the two face categories together with the original 1000 categories. We built an ensemble of 14 fine-tuned CNNs comprising Inception, ResNet \citep{he2016deep}, ResNeXt \citep{xie2017aggregated}, DenseNet \citep{huang2017densely}, and SENet \citep{hu2018squeeze}, and created adversarial images for the model ensemble using iterative gradient descent.

\begin{figure}[hpt]
\centering
\includegraphics{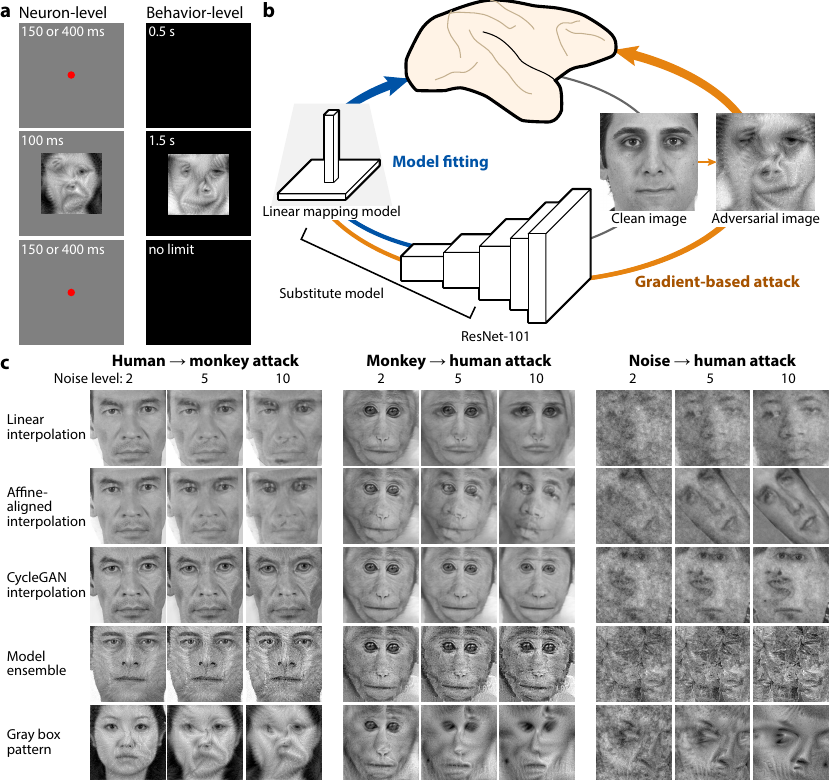}
\caption{Overview of image manipulations targeting categorization change. \textbf{a}, The left column illustrates monkey neuron-level experiments. Monkeys fixated on a red fixation point while images were presented in random order and neuronal responses were recorded. Images were presented for 100 ms; inter-stimulus interval was 150 ms for monkey 1 and 400 ms for monkey 2, whose responses extended over a longer duration. The right column illustrates behavioral experiments with human subjects. Each image was presented for 1.5 s. Humans were instructed to categorize the image by pressing a key. There was no time limit for a response. The text in the figure was not included in the experiment. \textbf{b}, A substitute model was fit on IT neuron responses and used to generate gray box deceptive images. The substitute model consisted of a pre-trained ResNet-101 (excluding the final fully-connected layer) and a linear mapping model. Deceptive images were generated by gradient-based optimization of the image to create a desired neuronal response pattern as predicted by the substitute model. \textbf{c}. Example images for human→monkey deceptive images (left), monkey→human deceptive images (center), and brown noise→human face deceptive images (right) are shown for different noise levels and different image manipulation methods.}
\label{fig:overview}
\end{figure}

Lastly, we designed a manipulation method tailored to primate vision by building a model of macaque visual neuron responses as a substitute model for perceptual change (\textbf{Figure~\ref{fig:overview}a}). The model comprised a ResNet-101 from the fine-tuned CNN ensemble above, fitted with a linear mapping module that used features extracted from the last convolutional layer to predict neuronal responses. Similar CNN-based models comprise the current best models of primate visual neuronal responses \cite{schrimpf2018brain}. The linear mapping was trained on responses of 22 face patch neurons in one monkey (of two in this study) to around 1,000 pictures of objects \citep{konkle2010conceptual}. The model could explain around 40\% of the neuronal response variance on held-out images not used during fitting. We used this neuron-fitted model, which was end-to-end differentiable, to stand for the primate visual system for adversarial attack. To create deceptive images, we used two variants of objective functions and two variants of optimization algorithms, resulting in four total variants. The objective function was either 1) to maximize responses in one model neuron that produced monkey-like features in feature visualization, labelled the \emph{single-neuron} method; or, 2) to match the model-predicted response pattern to the empirical neuron population response pattern to monkey faces, labelled the \emph{pattern} method. The objective function was optimized using either 1) iterative gradient descent, or 2) iterative gradient descent coupled with $l_2$-projection (\emph{$l_2$-PGD}), i.e., projection to a fixed noise level (MSE, equivalent to $l_2$) at each step of gradient descent. We refer to these variants collectively as \emph{gray box} deceptive images, as they use a limited amount of information about the system (primate vision) being targeted.

Each method was used to modify 40 unmodified or \emph{clean} human faces, except the single-neuron method, the pattern method, and their $l_2$-PGD versions, which were each used to modify 20 images. Each method produced one deceptive version of each clean image at every noise level. Thus, we tested a total of 2,400 targeted human→monkey deceptive images.

\begin{figure}[hpt]
\thisfloatpagestyle{empty}
\centering
\hspace*{-0.25in}
\includegraphics{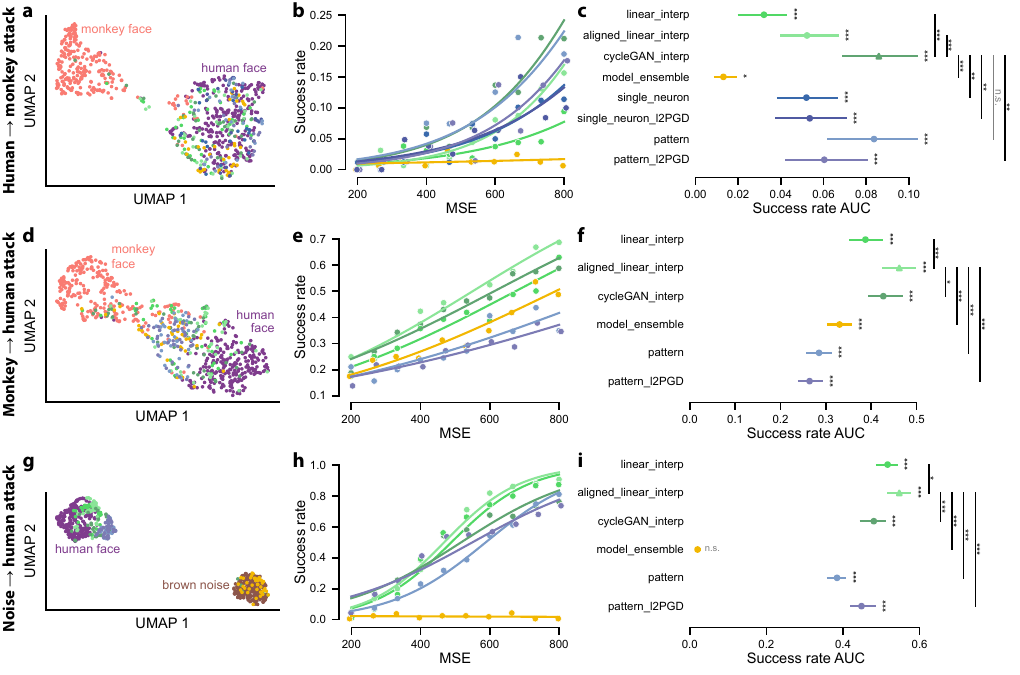}
\caption{Responses to deceptive images by face-selective neurons in monkey IT. \textbf{a--c} corresponds to human→monkey deceptive images. \textbf{a}, The scatter plot shows a UMAP visualization of the neuron population representation of images, showing clean human faces, monkey faces, and level 10 deceptive images.  The neuron pseudopopulation included IT neurons from the face patches of two monkeys. The colors indicate image manipulation methods and are labeled in \textbf{c}. \textbf{b}, The plot shows success rate as a function of noise level (MSE). The scatter points represent method-level (image-averaged) success rate per noise level. The lines represent logistic regression fit on image-level success rates. \textbf{c}, The plot shows success rate AUC per manipulation method. Each point and line indicate the mean and bootstrap 95\%-CI of the mean, respectively. The triangle marker signifies the method with the maximum AUC. The annotations on the right indicate results of pairwise statistical tests for a difference in AUC. The annotations next to individual methods indicate whether success rate increased with noise level using a permutation test. *, p < 0.05; **, p < 0.01; ***, p < 0.001; and n.s., not significant, all p-values FDR-corrected for multiple comparisons.  \textbf{d--f} are the same as \textbf{a--c} but correspond to monkey→human deceptive images. \textbf{g--i} are the same as \textbf{a--c} but correspond to brown noise→human deceptive images.}
\label{fig:neural}
\end{figure}

We presented the deceptive images, together with 250 unmodified human faces and 250 monkey faces (\textbf{Figures~\ref{fig:image_examples}}), to two monkeys in a passive fixation task and recorded neuronal responses using chronically implanted multi-electrode arrays. Specifically, we recorded from face patches in inferior temporal cortex (IT), which contain neurons that are strongly selective for faces over non-face objects and that extract face features including species and identity \citep{tsao2006cortical,moeller2008patches}. To collect enough repeat presentations to reliably measure neuron responses, deceptive images were presented in 4 sessions of 2--3 noise levels each.

We visualized the neuron population representation of clean and deceptive images (level 10) using Uniform Manifold Approximation and Projection (UMAP) \citep{mcinnes2018umap} (\textbf{Figure~\ref{fig:neural}a}). Neuronal representations of clean human and monkey faces were clearly separable. Some deceptive images were shifted away from human faces towards monkey faces. To quantify what fraction of the deceptive images would be categorized as the target category, we trained support vector machines (SVMs) to classify clean images as human or monkey based on neuronal responses. The SVMs were then used to classify held-out clean images as well as the deceptive images. The SVMs achieved high test accuracy on held-out clean human and monkey faces (98.3 ± 0.3\% and 97.7 ± 0.5\%, mean ± sem; \textbf{Figure~\ref{fig:neural_sup1}c}), confirming that the recorded neurons distinctly represented human and monkey faces. Nevertheless, a fraction of human→monkey deceptive images were classified as monkey faces. We call these images \emph{successful}, and calculated \emph{success rate} as the fraction of successful images per method at each noise level (\textbf{Figure~\ref{fig:neural_sup1}a}). The success rate of almost all manipulation methods increased with noise level (\textbf{Figure~\ref{fig:neural}b}), as reflected in the positive coefficient in logistic regression of individual image success as a function of noise level (\textbf{Figure~\ref{fig:neural}c}). The best manipulation method was CycleGAN-based interpolation, which reached a success rate of 21 ± 4\% (mean ± sem) at noise level 10, followed by the gray box pattern method (19 ± 6\%), pattern $l_2$-PGD (18 ± 5\%), and affine-aligned linear interpolation (16 ± 4\%; \textbf{Figure~\ref{fig:neural_sup1}c}). To summarize and compare across methods, we calculated the area under the success rate curve (AUC) using the logistic fits (\textbf{Figure~\ref{fig:neural_sup1}c}). We normalizing the x-value range to 0--1, so that the AUC value is roughly equivalent to the average success rate over levels. The best method, CycleGAN-based interpolation, was significantly more successful than all other methods except the gray box pattern method (p = 0.068; permutation test, FDR-corrected across 7 tests; \textbf{Figure~\ref{fig:neural}c}).

Logistic regression also allowed us to infer how much image change would be needed to achieve 50\% successful deception with each method (\textbf{Figure~\ref{fig:neural_sup1}b}). Since no method achieved 50\% success rate within the tested noise levels, this midpoint estimate is necessarily an extrapolation. Nevertheless, an estimated midpoint may be useful to compare to other perceptual change directions (below) and to the much larger MSE separating clean images. CycleGAN interpolation would require MSE = 1030 (bootstrap 95\%-CI: 930--1150) to achieve 50\% success rate.  Simply replacing a human face with the closest monkey face among 250 images, while guaranteeing complete success, requires MSE = 3210 (95\%-CI: 1670--4490; this differs from 2800 reported above because the clean images subjected to manipulation are different from the 250 images used for establishing a baseline). This closest distance increases to 4560 and 6800 for closest among 25 and 3 images respectively, a roughly 1.5-fold increase for each order of magnitude; extrapolating in the other direction suggests that even if the trend continues, there may only be one in over 10\textsuperscript{5} monkey face images that is MSE ≈ 800 away from an average human face image. Interpolating halfway between a human face and the closest monkey face among 250 entails MSE = 800 (95\%-CI: 420--1120), but that likely corresponds to lower-than-50\% success rate because empirically, linear interpolation at MSE = 800 achieved only 10 ± 4\% success rate. Overall, these comparisons suggest that the semantic change success achieved by CycleGAN-based interpolation and the pattern method was not explained by trivially replacing the original image. 

Were deceptive images being misclassified simply because noise degraded image quality? To control for this possibility, we tested two types of control image modifications at noise level 10. The first was image degradation, including Gaussian noise, Gaussian blur, phase scrambling, and Eidolon images at three coherence levels \cite{koenderink2017eidolons,tramer2020fundamental}. The second was versions of targeted deceptive images where the image change (additive pixel value change) was re-applied flipped upside-down \cite{elsayed2018adversarial}.

Considered together, deceptive images achieved significantly higher success rate than either clean human face images or non-targeted image modifications (p = 1 × 10\textsuperscript{-4}; permutation test, FDR-corrected across 2 tests; \textbf{Figure~\ref{fig:neural_sup1}c}). Considering each method individually, among control images, flipped CycleGAN images achieved the highest success rate (16 ± 3\%), a success rate that was statistically no lower than success in all targeted manipulation methods (all pairwise p > 0.13; permutation test, uncorrected). This notwithstanding, the next best among 14 control methods, the flipped single neuron method, achieved only 4 ± 3\% success rate. Thus, we speculate that flipped CycleGAN images still contain features of a monkey face to which neurons are sensitive. To anticipate, results below from human categorization support the conclusion that flipped CycleGAN images were not perceived as monkey faces.

\subsection*{Deceptive Images in Two Other Directions Also Led to Target Category-like Responses}
So far, we described human faces modified to look like monkey faces (\emph{human → monkey} or \emph{h2m} manipulation). In adversarial attack in CNNs, one category can be attacked to target any other category~\cite{szegedy2013intriguing}. Therefore, we attempted to create deceptive images in two other directions: the reverse \emph{monkey→human} (\emph{m2h}) direction and the \emph{brown noise→human face} (\emph{b2h} or \emph{noise→human}) direction, a shift between two distant categories. We tested these directions with the same methods, except that for the gray box method, we only tested the match-response-\emph{pattern} objective and not the excite-\emph{single-neuron} objective.

Unexpectedly, monkey→human manipulation on neuronal responses was generally easier than human→monkey direction, as reflected by overall higher success rates for most methods at most noise levels (\textbf{Figure~\ref{fig:neural}e,f}) and by the highest success rate achieved at noise level 10 (69 ± 34\% for affine-aligned linear interpolation; \textbf{Figure~\ref{fig:neural_sup1}f}). UMAP visualization (\textbf{Figure~\ref{fig:neural}d}) shows that, unlike in human→monkey manipulation, many monkey→human deceptive images had the target category (clean human faces) as the majority of near neighbors. However, perceptual change success was also high for control image modifications, including untargeted image degradation (\textbf{Figure~\ref{fig:neural_sup1}f}). This cannot be because neurons did not separate the categories of monkey and human faces; the categories were clearly separated, as we established above in human→monkey direction and further verified with the particular experimental sessions here (accuracy on human and monkey faces: 98 ± 5\% and 98 ± 8\%; \textbf{Figure~\ref{fig:neural_sup1}f}). Instead, we speculate that the idiosyncrasies of the neuronal selectivity or the particular monkey images used in the manipulation could have prevented SVM generalization. Failing to certify the SVM-based quantification, we could not meaningfully compare different manipulation methods or estimate the noise threshold for successful perceptual shift. We will show below that human behavior distinguished effective targeted deceptive images from control image modifications.

Further counter to our prediction, brown noise→human face manipulation was still easier than the previous two directions. The most successful method, as quantified by AUC, was affine-aligned linear interpolation, which was significantly better than all other targeted methods (p ≤ 0.014; \textbf{Figure~\ref{fig:neural}i}) and reached 91 ± 17\% success rate at level 10, approaching the accuracy on clean human face images (98 ± 5\%; \textbf{Figure~\ref{fig:neural_sup1}i}). Linear interpolation (level 10 success rate: 87 ± 20\%) and CycleGAN based-interpolation (76 ± 34\%) were also highly effective, followed by gray box manipulation methods (pattern: 81 ± 24\%; pattern $l_2$-PGD: 74 ± 25\%). The model ensemble method was ineffective, achieving 0.6 ± 4\% success rate at level 10, close to the error rate on brown noise images (0.4 ± 2\%). UMAP visualization (\textbf{Figure~\ref{fig:neural}g}) qualitatively corroborates the quantification, with most interpolation-based images intermixing with clean human faces, gray box images close to but forming a distinct cluster from human faces, and model ensemble images remaining embedded in the noise images cluster. The most successful control image modification (Gaussian blur) achieved only 16 ± 4\% success at level 10, significantly lower than success rates of all targeted methods (all p = 2 × 10\textsuperscript{-5}; permutation test, FDR-corrected across 6 tests) excluding the model ensemble method (p = 0.175), indicating that targeted manipulation methods were specific (\textbf{Figure~\ref{fig:neural_sup1}i}). Using fitted logistic regression, the estimated noise threshold for 50\% successful perceptual change by the best method (affine-aligned interpolation) was MSE = 470 (95\%-CI: 450--490), much lower than that for simply replacing the noise image with the closest human face (MSE = 1850, 95\%-CI: 1530--2220), although close to that for interpolating half way between the two (MSE = 460, 95\%-CI: 380--550). \textit{Post hoc}, we could rationalize the relative ease of noise→human semantic flip by suggesting that a face superimposed on noise can still look like a face. This may be attributed to the heightened sensitivity of primates in detecting faces (pareidolia), or the fact that noise images do not themselves belong to any category that can supply competing evidence to the positive evidence of a face. 

To summarize, we tested changing image categorization by monkey neuronal responses through targeted image perturbation in three directions. Within the range of image change tested, we achieved moderate success in one direction (human→monkey) and high-to-complete success in the other two (monkey→human and brown noise→human face). In human→monkey manipulation, although the highest success rate was lower than 50\%, the success was not explained by the control method of interpolating the original image toward the target. In noise→human manipulation, close to complete success was achieved by control interpolation methods. We could not interpret the results in monkey→human manipulation since most types of non-targeted image degradation also achieved high success. We conclude that the ease of  deceptive images for monkey neurons was highly dependent on the involved categories and shift direction. In all manipulation directions, deceptive images tailored to CNNs (the model ensemble method) did not affect monkey neuronal representation, consistent with the general intuition that CNN adversarial images do not affect biological vision. There was no consistent evidence that the model ensemble method was more successful in the initial part of the response in a time-resolved analysis (\textbf{Figure~\ref{fig:neural_sup2}}).

\subsection*{Deceptive Images Changed Human Categorization}

\begin{figure}[hpt]
\centering
\includegraphics{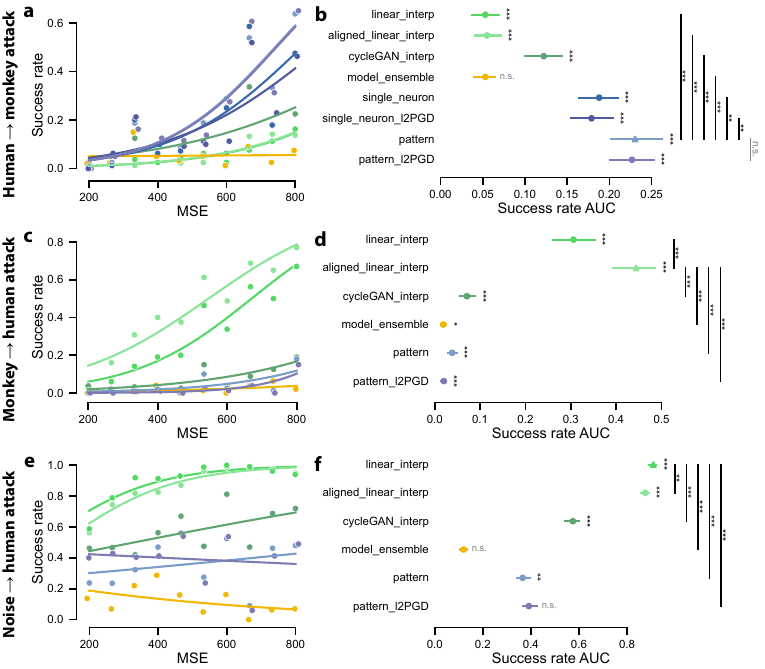}
\caption{Results of image targeted manipulation on human behavior in a visual categorization task. Format of the plot follows that in Figure \ref{fig:neural}. Panels \textbf{a, b} correspond to human→monkey deceptive images; \textbf{c, d} correspond to monkey→human deceptive images; \textbf{e, f} correspond to brown noise→human deceptive images.}
\label{fig:mturk}
\end{figure}

Could the same deceptive images also mislead human judgment? We recruited human subjects on Amazon Mechanical Turk (MTurk) to categorize deceptive images in a two-way categorization task. Deceptive images were shown intermixed with an equal number of clean images, a design we chose to encourage the subjects to make their best attempt at categorizing every image and to provide us with a way to monitor performance. Subjects had ample time to examine the images (1 s, no backward masking) and to make the choice (up to 4 s). For each deceptive image, we collected responses from 4--5 subjects who completed the task and had high performance on clean images. Each subject was tested on only one noise level of deceptive images.

Among human→monkey deceptive images, the four variants of gray box manipulation were the most effective methods. At noise level 10, these methods achieved up to 46--65\% success rate (\textbf{Figure~\ref{fig:mturk}a}). The best variant was the pattern method, with 64 ± 6\% success rate at noise level 10 and a success rate AUC that was higher than for all other methods, in particular linear interpolation and affine-aligned linear interpolation (both p = 3 × 10\textsuperscript{-5}; permutation test, FDR-corrected across 7 tests).

In the monkey→human direction, affine-aligned linear interpolation was the most successful method, achieving 77 ± 29\% success rate at noise level 10. In the noise→human direction, the most successful method was linear interpolation (without affine-alignment) with 94 ± 9\% success at nose level 10. In this direction, success rate started at noise level 1 at around 60\% for the most effective methods and around 15\% for the least effective method (model ensemble). This high sensitivity may be due to the tendency of humans to see pareidolia faces combined with a priming effect in the instruction, `Determine whether each image is a face image or a non-face image.'

For human behavior in all manipulation directions, control image modifications always resulted in lower success rates than the best targeted manipulation methods (\textbf{Figure~\ref{fig:mturk_sup}c, f, i}), indicating that the targeted methods were specific and allowing us to estimate the threshold noise level for 50\% successful manipulation. The non-targeted modification that led to the highest success rate was Eidolon 3 (i.e., Eidolon with coherence = 0; success rate 14 ± 5\%) for human→monkey direction, Gaussian blur (success rate 31 ± 21\%) for monkey→human direction, and flipped linear interpolation (success rate 18 ± 4\%) for noise→human direction. The threshold noise level for the most successful manipulation method in each direction was MSE = 750 (95\%-CI: 710--790) for human→monkey direction, 550 (95\%-CI: 510--600) for monkey→human direction, and 70 (95\%-CI: 10--140; outside the tested range of noise and thus an extrapolation) for noise→human direction.

To summarize, we changed human visual categorization through minimum targeted image manipulation, achieving > 50\% success in all three manipulation directions. As in monkey neuron-level results, we found a better manipulation method than interpolation-based control methods only in the most difficult human→monkey direction. Here, the gray box pattern approach was the best method, while CycleGAN-based interpolation was much less effective. As in monkey neurons, some manipulation directions were easier than others, with noise→human direction being the easiest, followed by monkey→human and lastly human→monkey. Finally, as in monkey neurons, CNN-tailored adversarial images were generally ineffective.

\subsection*{Deceptive Image Success Was Correlated between Monkey Neuron Responses and Human Behavior}

\begin{figure}[hpt]
\centering
\includegraphics{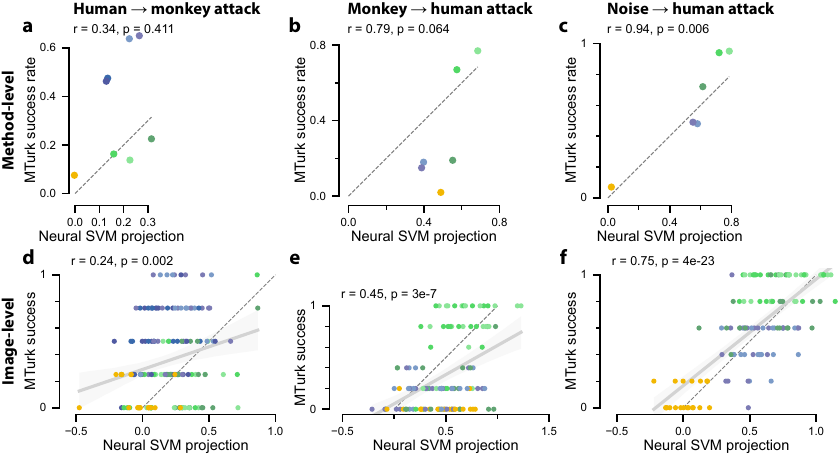}
\caption{The success of deceptive images was correlated between monkey neuronal responses and human behavior. The y-axis shows the success of deceptive images on human behavior; the x-axis shows the success of deceptive images on monkey neuronal responses. The dotted lines indicate identity. The colors indicate image manipulation methods and are labeled in \textbf{Figure~\ref{fig:neural}}. \textbf{a--c}, The scatter plots illustrate method-level (image-averaged) success for deceptive images at noise level 10. \textbf{d--f}, The scatter points illustrate per-image success for deceptive images at noise level 10. The solid lines indicate linear regression of the points, because many points are occluded due to the discrete y-values. Panels \textbf{a, d} correspond to human→monkey deceptive images; \textbf{b, e} correspond to monkey→human deceptive images; \textbf{c, f} correspond to brown noise→human deceptive images.}
\label{fig:compare_hm}
\end{figure}

To more directly compare the results of perceptual change in monkey neurons and human behavior, we correlated image manipulation success across methods (image-averaged) or across individual images (subject-averaged). To capture a more graded measure of success at the neuron level, we used the projection of an image's trial-averaged response vector onto the norm of the decision hyperplane of linear SVMs. The projection value was normalized and shifted so that the mean location of source and target class images was 0 and 1 respectively. Thus, the projection value was usually but not always between 0 and 1. For human behavior, there was no equivalent image-level measure (each subject only saw each deceptive image in one trial, and chose either one or the other category), so we quantified the success of each image as the fraction of subjects who chose the target category.

Success was positively correlated in money neuron and human behavior results, with generally high correlation values (r > 0.34 on the method-level and r > 0.24 on the single image-level; \textbf{Figure~\ref{fig:compare_hm}}). Except on the method-level in human→monkey and monkey→human directions, the correlations were statistically significant (p < 0.006; exact test against null hypothesis that two samples are from independent Gaussian distributions, uncorrected). In fact, human-to-monkey correlation was as high as possible after taking into account the between-subject consistency (\textbf{Figure~\ref{fig:model}a--d}). These analyses were limited to level 10 images only. Because success rate generally increased with noise level, including images from all noise levels introduces a confounding factor into correlation values. We report the analysis results including all noise levels in \textbf{Figure~\ref{fig:compare_hm_sup}}.

\subsection*{Deceptive Images Reveal Mismatch between Primate Vision and ANNs}

\begin{figure}[hpt]
\centering
\thisfloatpagestyle{empty}
\includegraphics{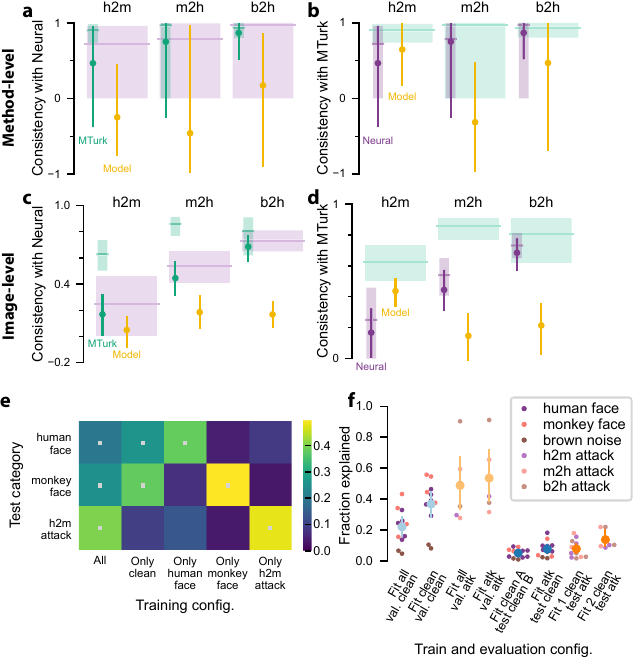}
\caption{Deceptive images reveal one aspect of mismatch between primate vision and ANNs. \textbf{a, b}, The plots illustrate the similarity, among three visual systems, in method-level (image-averaged) success of noise-level-10 deceptive images. We compared monkey neuron responses (`Neural'), human behavior (`MTurk'), and ANN model categorization (`Model'). The similarity in the pattern of deception success was quantified by Pearson's correlation across methods (\textbf{a, b}) or images (\textbf{c, d}). \textbf{a, c}, The success pattern in monkey neurons was compared to human and model. \textbf{b, d}, The success pattern in human was compared to monkey and model. The scatter points and vertical lines indicate correlation between visual systems (mean and bootstrap 95\%-CI of the mean). The horizontal lines and shaded areas indicate inter-subject split-half self-consistency (mean and CI). \textbf{e}, The heat map shows the performance of ANN-based predictive models of monkey neuronal responses, in conditions of interpolation (testing on held-out images from trained categories) and extrapolation (testing on images from held-out categories). Model performance was quantified by the fraction of response predicted (see Methods). Categories included in training are indicated by small grey squares in the heat map. \textbf{f}, results in \textbf{e} are summarized and combined over different deceptive image directions. Small dots indicate individual cell values as in \text{e}, color coded by the image category. Larger dots with whiskers indicate mean and bootstrap 95\%-CI of the mean within each training configuration, color coded by whether the result corresponds to interpolation (lighter color) or extrapolation (darker color) performance; and whether the tested category was clean (blue) or deceptive (orange) images.}
\label{fig:model}
\end{figure}

ANNs are expected to differ from primates in responding to adversarial attack, not least because ANNs should be highly sensitive to ANN-tailored adversarial images (the model ensemble method), which barely affected humans and monkeys. We measured ANN adversarial robustness in the same setting as the primate experiments, testing the same ensemble of 14 fine-tuned CNNs used to generate model ensemble deceptive images. Individual network outputs (logits) were averaged, then converted into a confidence vector over 1002 classes using the softmax function.

In human→monkey and monkey→human directions with the model ensemble method, success rate was already saturated at the lowest noise level 1 (\textbf{Figure~\ref{fig:model_sup}a, d}), an expected result since noise level 1 is already large relative to the typical threshold reported in machine learning literature. Testing additional images at lower noise levels, we verified that CNNs could be fooled with a minuscule amount of image change  (\textbf{Figure~\ref{fig:model_sup}b, e}), with 50\% success thresholds at MSE = 2.1 (95\%-CI: 2.0--2.2) and 4.3 (95\%-CI: 4.1--4.5) respectively for the two manipulation directions. Compared to the threshold for categorization change for primates as a minimum over methods, the threshold for CNN thresholds were over 100 times lower. Granted, this ratio is a likely overestimation because we could much more readily optimize deceptive image for CNNs than for primates. Nevertheless, this ratio is a direct, quantitative comparison of model and primate adversarial robustness.

Unexpectedly, the brown noise→human face direction for CNN models was difficult, requiring a manipulation threshold at MSE = 5310 (95\%-CI: 5290--5340; \textbf{Figure~\ref{fig:model_sup}h})---even higher than the budget for simply replacing the noise image with the closest human face (MSE = 1880; 95\%-CI: 1620--2210), an approach that would have led to perfect success (\textbf{Figure~\ref{fig:model_sup}h}). This shows that a simple gradient descent algorithm could not find the optimal deceptive image at such high noise levels. The criterion for noise→human shift success might be more stringent than in other directions, because the deception counted as successful only if confidence in the human category was higher than confidence in all other 1000 categories (excluding the monkey category). In contrast, in human→monkey manipulation for example, confidence in the human category only needed to exceed confidence in the monkey category. Most adversarial attack studies use the stricter one-vs-all criterion. However, to the best of our knowledge, we are not aware of studies that attempted adversarial attack starting from a noise image. Brown noise images may be particularly hard to attack (compared to Gaussian noise) in our preliminary tests.

How did CNNs respond to deceptive images derived from other manipulation methods? Gray box perturbation methods were the second-most effective group of methods in human→monkey and monkey→human direction, although not in noise→human direction (\textbf{Figure~\ref{fig:model_sup}a, d, g}). In the human→monkey direction, gray box methods were almost as effective as the model ensemble approach and achieved 80--100\% success rate starting from noise level 1. This result is unexpected, because although gray box methods used a ResNet as the feature-extraction backbone, the objective function was independent of weights in the classification layer of the CNN. Indeed, the gray box approach was not as effective as model ensemble manipulation in the monkey→human direction. In the noise→human direction, linear interpolation was the most effective method, followed by CycleGAN interpolation and model ensemble manipulation.

Thus, the overall pattern of manipulation method effectiveness was different between CNNs and primates. We directly compared CNNs to primates by correlating perceptual change success on the method- and individual image-levels as above (\textbf{Figure~\ref{fig:model}a--d}). At the method level, correlation between model and primate was negative in 3 out of 6 comparisons (monkey or human results, 3 manipulation directions). At the image level, model-primate correlation was always positive, but lower than primate-primate correlation in 5 out of 6 conditions (in the exception case, monkey-human correlation was upper-bounded by low internal consistency).

This divergence between CNN and primate behavior suggests that CNN internal representation may also be less similar to neuronal representation than generally thought. To quantify this, we used the well-established approach of fitting a linear readout from internal features of a CNN (pre-trained on ImageNet object categorization) to predict neuronal responses. Typically, CNN-based predictive models of neuronal responses are tested on \emph{interpolation} to a held-out random subset of images that, by construction, are (in expectation) identically distributed as the training images. Here, we tested holding out entire categories of images, such as deceptive images, to test model \emph{extrapolation} to images that may come from a different distribution. An example scheme is shown in \textbf{Figure~\ref{fig:model}e}. Using data involving human→monkey deceptive images, we held out 1--3 categories of images (human faces, monkey faces, and/or deceptive images), fitted a linear ridge regression model from features extracted from ResNet-101, and tested the model on interpolation (unseen images from categories seen during training) or extrapolation (images from categories unseen during training). The models consistently performed much better on held-out images from categories included in training than on images from completely held-out categories. The same analysis, performed and pooled over all three manipulation directions, is presented in \textbf{Figure~\ref{fig:model}f}. When testing on a held-out subset of image categories seen during training, model validation performance was relatively high (first four groups in \textbf{Figure~\ref{fig:model}f}). However, when fitting on a subset of categories and testing on entirely held-out categories, performance was much lower (second four groups). Unexpectedly, generalization was about as poor from one clean image category to another (fifth group) as it was from deceptive images to clean images (sixth group) and vice versa (last two groups). Thus, although deceptive images revealed behavioral differences between CNNs and primates, the lack of generalization in CNN-based models of neurons could be already revealed by different categories of clean images, which likely had different statistics because they came from different image datasets. This conclusion did not change when we systematically repeated the analysis in \textbf{Figure~\ref{fig:model}e,f} on eight ANN models ranging from classical (ResNet-101, DenseNet169, and AlexNet) to state-of-the-art (Visual Transformer, EfficientNet\cite{xie_noisystudent}, and CLIP\cite{radford_CLIP}) to biologically focused (CORnet\cite{KubiliusSchrimpf2019CORnet} and VOneNet\cite{dapello2020simulating}); and across four regression methods (ridge regression, partial least squares, principal component regression, and factor analysis followed by linear regression) in combination with the best tested model, the Visual Transformer (\textbf{Supplementary Figures~\ref{fig:generaliz1}} and \textbf{\ref{fig:generaliz2}}).

\section*{Discussion}
Perception does not merely mirror reality, as evinced by centuries of visual illusions and a longer history of art. Adversarial images could constitute a kind of visual illusion, namely minimally changed images that were originally unambiguous, yet come to be categorized differently. They would be distinct from phenomena such as pareidolia, which pertain to images intrinsically ambiguous or misleading. We were motivated to study deceptive images for the primate brain by the similarities between CNNs and primate vision and by the former's high sensitivity to adversarial attack. We found deceptive images that fooled humans in visual categorization without strict time limits. Further, the altered categorization is reflected in monkey category-selective neuronal responses.

Some of the deceptive images were created using a CNN-based model of visual neuron responses. An ideal model of the brain should be interpretable, have high predictive power, and be useful for controlling the brain (a special case of prediction). CNNs are the best current models of visual cortex insofar as CNNs excel at predicting neuronal responses \cite{yamins2014performance, cadena2019deep,schrimpf2018brain} and, conversely, at synthesizing images to `control' those responses \cite{bashivan2019neural,walker2019inception}. Nevertheless, it is a topic of some controversy whether CNNs are easy to interpret \cite{serre2019deep,richards2019deep}. Since CNNs are vulnerable to adversarial attack whereas primates are thought not to be susceptible, images related to adversarial attack provide an ideal testing ground for CNN models of the brain. Adversarial images for CNNs will likely drive neuronal responses that are challenging to for a CNN-based model to predict. Meanwhile, creating deceptive images for the brain represents a challenging target to which to guide neuronal responses. Using a CNN-based model of (monkey) visual neurons to target a response pattern (i.e., the gray box methods), we could create deceptive images that guided (human) categorization behavior in two directions (human→monkey and noise→human). Our results bridge the gap between previous studies that outline a link between visual neuronal activity and behavior \citep{sheinberg1997role,afraz2006microstimulation,rajalingham2019reversible} and studies that show that computational models of neurons can be used to predictively control their activity \citep{bashivan2019neural,walker2019inception}.

However, using a CNN-based model of neurons to create deceptive images was neither consistently successful, nor as successful as the model predicted, nor always better than a naive method of simply interpolating toward a target class image. Indeed, CNN-based models could not adequately predict neuron responses to deceptive images when trained only on clean images. Furthermore, changing primate vision required much higher noise thresholds than changing model categorization, although future work may discover better methods for deceiving primate vision that require less image change. The pattern of deceptive image success across methods and individual images was also less similar between models and primates than within and between primate species. These results reveal a shortfall between CNN models and primate vision that is larger than previous studies suggest, and point to directions and clues for building better models.

\section*{Methods}
\paragraph{Data and code availability}
All stimuli, data underlying figures, code for data analysis, and code for generating deceptive images will be made available with the publication at \url{https://github.com/yuanli2333/Adversarial-images-for-the-primate-brain}. Source data are provided with this paper.

\paragraph{Subjects}
One adult male macaca mulatta (10 kg; 13 years old) and one adult male macaca nemestrina (13 kg, 11 years old) were socially housed in standard quad cages on 12/12 hr light/dark cycles. Animals were implanted with custom-made titanium headposts before fixation training. After several weeks of fixation training, the animals underwent a second surgery for array implantation. Monkey 1 was implanted with chronic microelectrode arrays (MicroProbes, Gaithersburg, MD) targeting the medial lateral (ML) face patch. Monkey 2 was implanted with chronic brush microwire arrays targeting the anterior medial (AM) face patch. Array targets were localized using fMRI aligned to CT scans and, during surgery, using landmarks from the CT scans. Localization was confirmed after surgery by CT scans. Extracellular electrical signals were amplified and recorded using the Omniplex data acquisition system (Plexon, Dallas, TX). All surgeries were performed under general anesthesia using sterile technique. All procedures on non-human primates were approved by the Harvard Medical School Institutional Animal Care and Use Committee, and conformed to NIH guidelines provided in the Guide for the Care and Use of Laboratory Animals.

Human psychophysics experiments were conducted online on Amazon Mechanical Turk. All participants provided informed consent and received monetary compensation for participation in the experiments. All experiments were conducted according to protocols approved by the Institutional Review Board at Boston Children’s Hospital.

\paragraph{Clean images}
We collected human and monkey faces from the web and from photographs taken in the lab to combine into 250 images per category. Monkey face images with excessively oblique head directions were excluded. Brown noise images were generated using custom code. Briefly, we generated a frequency-power spectrum that decayed as $1/f^2$, assigned random phase per image, and generated pixel values via inverse Fourier transform. Pixel values were scaled and shifted to have a valid range.

Source images used for manipulation were kept separate from other clean images used for establishing baseline accuracy. Source human faces images were cropped versions from the Chicago Face Database \cite{ma2015chicago}. We used 40 different images for each of four main type of manipulation method (interpolation-based, model ensemble, gray box, and non-targeted image degradation), totalling 160 source images. Source monkey faces were a set of 40 images separate from the set of 250 clean images; all manipulation methods used this set of 40 images. Source brown noise images were also different for each main type of manipulation method and totalled 160 images separate from the 250 clean images.

\paragraph{Deceptive and control images}
In all, we tested 3 manipulation directions and 6--8 targeted methods at 10 noise levels in each direction (human→monkey manipulation included 8 methods; the other two directions included 6). We tested the following control methods: 6 methods for non-targeted image degradation, and versions of targeted images with the image manipulation (pattern of additive pixel changes) flipped. Control methods were tested only at the highest noise level 10. Each method comprised 40 images at each direction and noise level, except human→monkey gray box methods, which each comprised 20 images, making a total of 8640 deceptive images.

\paragraph{Generating deceptive images}
All deceptive images were generated starting with a clean image $x$ from the source category. Linear interpolation images were generated as 
$$x_\text{adv}=(1-\lambda)x + \lambda x_\text{target},$$
where $x_\text{target}$ is different for each method, and $\lambda$ is a parameter from 0 to 1 tuned to give a desired noise level. For ordinary linear interpolation, $x_\text{target}$ is the closest image from 250 clean images of the target class. For affine-aligned linear interpolation, $x_\text{target}$ is the closest target class image after an affine transformation that was optimized for each source-target image pair. Boundary mode of affine transformation (for filling uncovered pixels after transformation) was the better one of either `reflect' or `constant,' in the latter case with the constant fill value being an additional parameter to be optimized. Unconstrained optimization sometimes resulted in images that no longer resembled the target class, e.g., by being scaled up to become an essentially constant image. Thus, affine transformation was subject to the following constraints: scale between 3:4 and 4:3; shear and rotation between -45° and 45°; and shift between -50 and 50 pixels (out of 224). For CycleGAN-based interpolation, $x_\text{target}=\text{CycleGAN}_{\text{source→target}}(x)$. We trained two CycleGAN models \citep{zhu2017unpaired}, one for translation between human and monkey faces and the other for translation between human faces and brown noise images.

Tuning of the parameter $\lambda$ was done in Python using scipy \citep{2020SciPy-NMeth} function `scipy.optimize.minimize' with method = `Powell'. Finding optimal affine transformation parameters was a generally difficult problem, and we used the best parameters found by any of 7 optimization algorithms, including scipy.optimize.minimize options `L-BFGS-B,' `TNC,' `SLSQP,' `COBYLA,' `Powell,' and `trust-constr'; and an CMA-ES family optimization algorithm (Algorithm 2 in \cite{loshchilov2014computationally}) we custom-implemented.

Both model ensemble and gray box deceptive images were generated using iterative gradient descent (related to \cite{kurakin2016adversarial} but distinct), 
with different objective functions. Specifically, each deceptive image $x_\text{adv}$ was iteratively generated as
\begin{equation} \label{eqn:adv_gen}
    x^{t+1}_\text{adv}= x^t_\text{adv} + \epsilon\cdot \bm{g}/\left(\text{stdev}(\bm{g})+\xi\right),\quad\bm{g}=\nabla_x \mathcal{L}(\theta,x^t_\text{adv}),
\end{equation}
where $x^t_\text{adv}$ is the deceptive image at iteration $t$; $\bm{g}=\nabla_{x}\mathcal{L}(\cdot)$ is the gradient of the cost function $\mathcal{L}$; $1/\text{stdev}(\bm{g})$ keeps gradient values at roughly the same scale across loss functions; $\xi=10^{-8}$ prevents underflow; $\epsilon$ is a learning rate that controls the amount of change in each step; and $\theta$ represents the substitute model parameters. For example, in human→monkey manipulation, $x^0_\text{adv}=x$ is a clean human face. The cost function at step $t$ is
\begin{equation}
\mathcal{L}=\mathcal{L}_\text{target}+\mathcal{L}_v,\quad\mathcal{L}_v=\sum_{i,j}\left(\left| x_{i+1,j}-x_{i,j} \right |+\left | x_{i,j+1}-x_{i,j} \right|\right),
\end{equation}
where $\mathcal{L}_\text{target}$ is a method- and target-specific loss function to be specified below, $\mathcal{L}_v$ is the total variation loss, and $(i, j)$ indexes pixels of the image. The total variation loss prevents high frequency noise from dominating the generated image features.

The target-specific loss function for the gray box pattern method was $\mathcal{L}_\text{target}=\norm{M_\theta(x_\text{adv}^t) - P_\text{target}}^2_2$, where $M_\theta$ is the substitute model, $P_\text{target}$ is the mean population neuron response pattern averaged over target class images, and $\norm{\cdot}^2_2$ is the $l_2$-norm between model-predicted and target response patterns.

The target-specific loss function for the gray box single-neuron method was $\mathcal{L}_\text{target}=-\left[M_\theta(x_\text{adv}^t)\right]_3$, where $\left[\cdot\right]_3$ indicates indexing the predicted activity of the third neuron (out of 22), a neuron we found to produce monkey-like features using feature visualization \citep{mahendran2016visualizing}. 

For model ensemble manipulation, $\mathcal{L}_\text{target}=-\left[M_\theta(x_\text{adv}^t)\right]_i$, where $i$ is the index of the monkey class for human→monkey manipulation and the index of the human class for monkey→human and brown noise→human face manipulation.

For all images generated with iterative gradient descent, noise level was controlled as follows. Noise level generally increased with the number of iterations, so we kept the last image during iteration that still fell into an MSE bin for each noise level 1--10. The bins were centered on the pre-specified MSE for each noise level with width MSE = 66.7 (the spacing between noise levels). Final images were projected to the pre-specified central MSE value. Learning rate was 0.03 for the pattern method and 0.08 for single neuron manipulation and model ensemble manipulation. Learning rate was lower (higher) for model ensemble manipulation with smaller (larger) MSE range in \textbf{Figure~\ref{fig:model_sup}b,e.h}. As many iterations as necessary were run to produce the specified noise levels, and generally ranged between 1,000 to 42,000 iterations.

All images generated with iterative gradient descent used jittering (40 pixels, randomized per iteration) as an additional regularizer for reducing high-frequency noise.

In $l_2$-PGD, the image during optimization was projected to the pre-specified MSE level, every iteration after that MSE level was first reached. Iteration continued until loss no longer decreased for 500 iterations. Images at the best iteration were saved. For each source image, we successively generated lower to higher noise levels, using each lower noise level deceptive image to seed optimization for the next higher noise level.

\paragraph{Generating control images}
Gaussian noise images were generated as $x_\text{gauss} = \text{clip}(x + \lambda g_x)$, where $g_x$ was sampled from the standard normal distribution, $\text{clip}(\cdot)$ keeps pixel values in the valid range, and $\lambda$ is tuned to yield a desired noise level.

Gaussian blur images were generated as $x_\text{blur} = \text{Blur}(x, \sigma)$, where $\sigma$ was the blurring filter kernel size tuned to yield a desired noise level.

Phase-scrambled images were generated as $x_\text{phase} = \text{PhaseShift}(x, \sigma)$, where $\sigma$ was the width of phase shift tuned to yield a desired noise level. The phase shift function was adapted from code from Geirhos et al. \citep{geirhos2018generalisation,geirhos_2018}. Briefly, it applies random uniform phase shift of the given width to the phase component of the image Fourier transform, symmetrically for the positive and negative frequency components, while leaving unchanged the amplitude component.

Eidolon images were generated as $x_\text{phase} = \text{Eidolon}(x, \text{reach}, \text{coherence})$, where coherence = 1, 0.3, 0 corresponds to what we refer to as Eidolon 1, Eidolon 2, and Eidolon 3 respectively, and reach was tuned to give a desired noise level. The Eidolon function was adapted from code from Geirhos et al. \citep{geirhos2018generalisation,geirhos_2018}, in turn based on a Python implementation \citep{gestaltrevision_2016} of the original paper describing Eidolon images \citep{koenderink2017eidolons}.

Upside-down flipped images were generated as $x_\text{flip}=\text{clip}(x+\text{flipUD}(x_\text{adv}-x))$. Around 0.3\% to 5\% of pixel values were clipped, resulting in image change values slightly under MSE = 800 for noise level 10.

Tuning of parameter(s) to achieve the specified noise level was done by optimization in Python using scipy \citep{2020SciPy-NMeth} function `scipy.optimize.minimize,' as described above.

\paragraph{Substitute model}
To generate deceptive images, a substitute model was trained to predict IT neuronal responses. The model was fixed before generating and testing deceptive images. The substitute model comprised a pre-trained CNN (ResNet-101 \citep{he2016deep} trained on ImageNet) and a linear mapping model (\textbf{Figure~\ref{fig:overview}B}). Because the 1,000 pre-trained categories did not include our categories of interest (human face and monkey face), we collected around one thousand images for these two categories for fine-tuning the ResNet-101. Next, a linear model was trained to map extracted image features (layer conv5\_3, the last convolutional layer) to neuronal responses. The linear model was factorized in the spatial and feature dimensions \citep{klindt2017neural}. The spatial module was a convolutional kernel $W_s$. The feature module was a mixing pointwise convolution $W_t$, i.e., a weighted sum over the feature dimension. Both $W_s$ and $W_t$ were parameterized separately for each IT neuron. Thus, the response for neuron $n={1,\dots,22}$ to image $x$ was modeled as
\begin{equation}
\hat{y}_{n}=\sum (W^{n}_{s}\ast \text{ResNet}_\text{conv5\_3}(x))\cdot W^{n}_{t}+W^{n}_{d},
\end{equation}
where $\ast$ denotes the convolution operation and $W^{n}_{d}$ is a bias parameter. The parameters were jointly optimized to minimize a loss function $\mathcal{L}_{e}$ composed of the prediction error $\mathcal{L}_p$, an L2-regularization loss $\mathcal{L}_2$, and a spatial smoothness loss $\mathcal{L}_\text{laplace}$:

\begin{equation}
\mathcal{L}_p = \sqrt{\sum_{n}(\hat{y}_n-y_n)^2}
\end{equation}
\begin{equation}
\mathcal{L}_{2} = \lambda_s\sum \norm{W_{s}}_2 + \lambda_{t}\sum \norm{W_{t}}_2
\end{equation}
\begin{equation}
\mathcal{L}_\text{laplace} = \lambda_l\sum W_s \ast \begin{bmatrix}
0 &-1  &0 \\
-1&4 &-1 \\
 0&-1&0
\end{bmatrix}
\end{equation}
\begin{equation}
\mathcal{L}_e = \mathcal{L}_p + \mathcal{L}_2 + \mathcal{L}_\text{laplace}
\end{equation}
The hyper-parameters $\{\lambda_s,\lambda_t,\lambda_l\}=\{1, 0.1, 0.7\}$ were obtained from a grid-search to produce the highest prediction accuracy. The substitute model achieved an average correlation of 0.4 between predicted and actual IT neuron responses on held-out test images.

\paragraph{Neuron-level experiment} We recorded neuronal responses to image presentation while monkeys did a passive fixation task. Images were presented on an LCD monitor (ASUS VG248, 165 Hz) at a viewing distance of 57--61 cm. Monkeys were required to fixate within a window of 1.5--3 degrees in exchange for juice reward. Eye position was monitored using an infrared eye tracker (EyeLink, Ottawa, Canada). Images were presented at 4 degrees of visual angle in size in randomized order in the receptive fields of the neurons. Each image was presented for 100 ms, with a 150 ms interval between images for monkey 1 and 400 ms for monkey 2 due to longer response dynamics. Image onset time and spike time stamps were aligned using digital event words and analog photodiode signal.

Neuronal responses to an image were calculated as firing rates averaged over trials within a response window after stimulus onset that was automatically selected per session to maximize split-half self consistency. The response window for monkey 1 started between 60--90 ms and ended between 240--280 ms; and for monkey 2, started between 125--180 ms and ended between 340--400 ms. Visually selective neurons were selected based on split-half self consistency > 0.1. Although self consistency is a function of number of trials, the cutoff of correlation = 0.1 is generous in our experience and compared to prior work \citep{bashivan2019neural}. Qualitative results were not affected when we repeated the analysis with different selection criteria. Responses per neuron were standardized to have zero mean and unit variance over images before downstream analysis.


\paragraph{Behavior-level experiment} Subjects on Amazon Mechanical Turk were invited to perform an image categorization task. They were instructed to determine `whether each image is a human face or a monkey face' for human→monkey and monkey→human manipulation sessions (subjects were not informed of this), or `whether each image is a face image or a non-face image' for brown noise→human face deception. Subjects were instructed to `Make your best guess. Sometimes, it may be hard to determine the correct answer.' To answer, subjects pressed the left arrow key for `Human face' or 'face,' and right arrow key for `non-face' or 'Monkey face,' as appropriate. Each image was presented for 1.5 s. There was no time limit for a response. No feedback was provided. Deceptive images were randomly and evenly intermixed with clean images, which allowed us to monitor performance. We selected subjects with > 95\% accuracy to include in further analyses. Responses were collected from 4--5 subjects for half of all deceptive images (20 per method per direction/noise level).

\section*{Quantification and statistics}
\paragraph{Quantifying size of image change}
Size of image change (noise level) was quantified by Mean Squared Error (MSE) as
$$\text{MSE}=\frac{1}{N}\sum_{i,j}\left(x'_{i,j}-x_{i,j}\right)^2,$$
where $x'$ and $x$ are a pair of images, and $i,j$ index $N$ total pixels. All images were standardized to be $N=224\times224$ pixels one-channel images with pixel value in the range 0--255. At this image size, MSE is related to $l_2$ (pixel value range 0--1) as 
$$l_2 = \sqrt{\sum_{i,j}\left(1/255\cdot\left(x'_{i,j}-x_{i,j}\right)\right)^2} = \sqrt{N/255^2\cdot\text{MSE}} \approx 0.8784\sqrt{\text{MSE}}.$$

\paragraph{Quantifying neuron- and behavior-level deception success}
Neuron-level deception success was quantified using linear Support Vector Machines (SVMs) separately for each experimental session. SVM fitting was carried out using the Python package `scikit-learn'~\citep{pedregosa2011scikit}. SVMs were trained to categorize clean images (two-way categorization) based on the corresponding neuronal responses. For example, in monkey→human perceptual change direction, the SVM was trained to classify each image as either a monkey face or a human face. We trained SVMs with balanced samples for 250 train-validation splits, leaving two images out in each split (one from each class). Moreover, we trained both linear and radial-basis function (RBF) SVMs, resulting in 500 classifiers per experimental session. The trained SVM was used to classify the held-out clean images, deceptive images, and control images. Perceptual shift success for each image was the fraction of classifiers that classified that image as the target class, a value that was between 0--1 but usually close to either 0 or 1. This value was further averaged across sessions and across 2 monkeys.

SVM projection was only defined for linear SVMs. It was calculated as the dot product between an image's response pattern and SVM classification weights, then shifted and scaled so that the category-average was 0 and 1 for the source and target categories, respectively.

Behavior-level manipulation success was quantified, for each image, as the fraction of trials/subjects in the target category was chosen  (each subject only saw each image once). 

Success rates for a method at a given noise level were calculated as the average success over images. Logistic regression was fitted on individual image-level data to capture image-level success as a function of MSE for each image (MSE was tightly distributed per noise level but not identical among images). Logistic regression was carried out using the Python package `statsmodels' \citep{seabold2010statsmodels}.

\paragraph{Time-resolved neuron responses analysis}
Plotted responses were smoothed with a uniform filter of width 25 ms. Principal components (`eigen-neurons') were derived from standardized neuronal responses as described above, then used to project the smoothed population response time courses. D-prime (for selecting example neurons) was calculated as $(\mu_1-\mu_2)/(\frac{1}{2}(\sigma_1 + \sigma_2))$, where $\mu$ and $\sigma$ indicate mean and standard deviations of responses to one image category. SVMs fitting was done as described above, repeatedly for successive, non-overlapping 10 ms responses windows.

\paragraph{Model prediction of neuronal responses in Figure \ref{fig:model}}
Instead of using the method described above to build a substitute model, for the analysis in \textbf{Figure~\ref{fig:model}}, we used a simpler and much faster modeling approach following Zhuang et al \citep{zhuang2021unsupervised}, which is in turn similar to prior work \citep{yamins2014performance,schrimpf2018brain}. Neuronal response models were based on features extracted by an ImageNet-trained ResNet-101 and fitted using partial least squares (PLS) regression with 5 retained components. We searched for the optimal ResNet layer (out of 105 including raw pixels) to use per monkey/array, using data not included in the results. We chose layer `layer3.5.conv1' for monkey 1 and `layer3.6.conv1' for monkey 2. For each category we used images. Training was done using 5-fold cross-validation, i.e., holding out each 20\% of the data in turn for testing. Cross validation was stratified; for example, when fitting on all image categories, each category had almost the same number (difference no larger than 1) of images in each training/validation fold. All reported model performance values were based on held-out images. PLS regression and stratified cross-validation was implemented by the Python package `scikit-learn' \citep{pedregosa2011scikit}.

Fraction of neuronal responses explained by model was quantified similar to prior work \citep{bashivan2019neural,schrimpf2018brain,zhuang2021unsupervised}. Specifically, we calculated the fraction explained per neuron as the square of a ratio whose nominator was the Pearson correlation between model prediction and neuron responses, and whose denominator was the split-half self consistency of that neuron (Pearson's r, Spearman-Brown corrected). This value resembles fraction of variance explained, i.e., coefficient of determination or $R^2$, although they are not equivalent unless an additional optimal linear transform was fit between the predicted and actual neuronal responses on the test set to remove any mean and scale differences. Fraction of variance explained, or $R^2$, was not suited for describing model performance on held-out data; $R^2$ was not directly used in the cited prior studies either. To produce one value for each training-testing configuration for each data set, we took the average over splits and median over neurons. 

\paragraph{Model prediction of neuronal responses in Supplementary Figures \ref{fig:generaliz1} and\ref{fig:generaliz2}} To verify that the results of Figure \ref{fig:model} were not specific to one ANN model and one fitting method we tested, we extended the analysis to systematically test eight ANN models and four regression methods. Using ridge regression, we tested three ``classical' CNN models (ResNet-101, DenseNet169, and AlexNet), three state-of-the-art ANN models (Visual Transformer, EfficientNet\cite{xie_noisystudent}, and CLIP\cite{radford_CLIP}), and two biologically inspired models (CORnet\cite{KubiliusSchrimpf2019CORnet} and VOneNet\cite{dapello2020simulating}). Table \ref{table:layers} reports the layers used for each model, chosen based on independent (not used for training or testing) recording sessions from the same monkeys. Additionally, we tested the Visual Transformer model in combination with four regression methods: ridge regression (alpha = 2 × 10\textsuperscript{5}), partial least squares (PLS, number of components = 27), principal component regression (PCR, number of components = 200), and factor analysis followed by linear regression (Factor, number of components = 250). For all regression methods, we used the Python package `scikit-learn'.

\begin{table}[h!]
\hspace{-0.35in}
\begin{tabular}{|c | c | c|} 
 \hline
 Model & Layer & Pretrained model \\ [0.5ex] 
 \hline\hline
 ResNet-101 & \begin{tabular}{@{}c@{}}layer3.5.conv1 (monkey 1), \\ layer3.6.conv1 (monkey 2)\end{tabular}   & torchvision models \\ \hline
 DenseNet169 &  features.norm5 & torchvision models   \\ \hline
 AlexNet & features.10 &  torchvision models   \\ \hline
 Visual Transformer & blocks.10.norm2 &  Python `timm' package  \\ \hline
 EfficientNet & blocks.3.0.conv\_dw &  Python `timm' package  \\ \hline
 CLIP & visual.transformer.resblocks.5.mlp.c\_proj & Github repo for Radford et al., 2021\cite{radford_CLIP}  \\ \hline
 CORnet &  V4.conv\_input &  Github repo for Kubilius et al., 2019\cite{KubiliusSchrimpf2019CORnet} \\ \hline
 VOneNet & module.model.layer2.3.conv2 & Github repo for Dapello et al., 2020\cite{dapello2020simulating}  \\ [1ex] 
 \hline
\end{tabular}
\caption{Layers from each ANN model used to build predictive models of neuronal responses, and the implementation source for each model.}
\label{table:layers}
\end{table}

\paragraph{Consistency between and across readouts in \textbf{Figures~\ref{fig:compare_hm} and \ref{fig:model}}}
Consistency was calculated as Pearson's correlation coefficient. Consistency within a readout was calculated by splitting the data in half by subjects, then correlating across method- (\textbf{Figure~\ref{fig:model}}a,b) or image-averages (\textbf{Figure~\ref{fig:model}c,d}), clipped at 0, then corrected using the Spearman-Brown formula.

\paragraph{Center and spread estimates}
Center estimates of AUC and deception success threshold MSE were based on the logistic fit to the original data; confidence interval (CI) estimates were based on bootstrapping 500 times over images and fitting a logistic function to each bootstrap sample. Center and CI estimates of success rate at one noise level were based on original data and bootstrapping over images at that noise level. Center and CI estimates of the threshold MSE of `Replace,' `Interp. half way,' and `Erase' was based on original data and bootstrapping over the set of source images. Center and CI estimates of consistency within and between readouts (\textbf{Figure~\ref{fig:model}}) were based on original data and bootstrapping over methods or images as appropriate.  

\paragraph{One-sample statistical tests for positive slope in logistic regression}
Logistic regression was repeated with data in which noise level (MSE values) was permuted relative to deception success. Permutation was repeated 10,000 times to build a null distribution of the model `slope' parameter. One-tailed p-value was calculated as the fraction of null distribution parameters that was larger than the sample value. P-values were clipped at a minimum of 1/10,000.

\paragraph{Two-sample statistical tests for difference between methods or method groups}
For each pair of groups/conditions compared, the group label in data was permuted relative to deception success (and noise level if applicable). For tests comparing AUC between methods, logistic regression was performed with permuted data. This creates a null distribution of pairwise difference in AUC, to which the sample value was compared. For tests comparing success rate at one noise level, the sample success rate difference was compared to the permuted null distribution of success rate differences. One-tailed p-value was the fraction of null values that was larger than the sample value. P-values were clipped at a minimum of 1/10,000.

\paragraph{Correction for multiple comparisons}
Correction was performed over each group of related tests (indicated in text) to control false discovery rate at the level of 0.05, using the two-stage Benjamini-Krieger-Yekutieli procedure \citep{benjamini2006adaptive} as implemented in the Python library `statsmodels' \citep{seabold2010statsmodels}.

\section*{Acknowledgments}
This work was supported by NIH grants R01EY16187, P30EY012196, R01EY026025, and NSF STC award CCF-1231216 to the Center for Brains, Minds and Machines at MIT. Jiashi Feng was supported by grants AI.SG R-263-000-D97-490 and MOE Tier-II R-263-000-D17-112. We thank Hu Zhang, David Castro, Ariana Sherdil and Peter F. Schade for discussion and assistance during the work.

\section*{Author contributions}
JF, WX, and LY conceived of the study. GK, MSL, WX, and LY designed the experiments. LY developed the code for creating adversarial images. MSL and WX acquired the neuronal recording data. LY acquired the human behavior data. LY and WX analyzed the data and drafted the manuscript. GD performed the systematic testing of ANN architectures and fitting methods for building neuronal encoding models. All authors interpreted the data and revised the manuscript. GK, MSL, FEHT, and JF provided funding.

\section*{Competing interests}
The authors declare no competing interests.

\clearpage
\baselineskip12pt
\bibliography{ref}
\clearpage
\nolinenumbers

\section*{Extended data figures}
\setcounter{figure}{0}
\makeatletter
\renewcommand{\thefigure}{S\@arabic\c@figure}
\makeatother

\begin{figure}[hpt]
\centering
\includegraphics{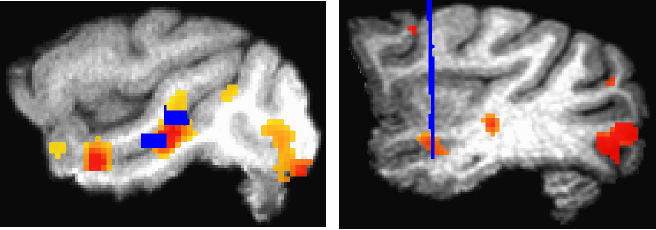}
\caption{Related to \textbf{Figure~\ref{fig:overview}}. Location of recording arrays in both monkeys is shown in relation to fMRI-defined patches selective for faces over objects. Blue indicates arrays localized by CT; red indicates face selectivity (thresholded at false discovery rate = 0.01).}
\label{fig:arrays}
\end{figure}

\begin{figure}[hpt]
\thisfloatpagestyle{empty}
\centering
\hspace*{-0.5in}
\includegraphics{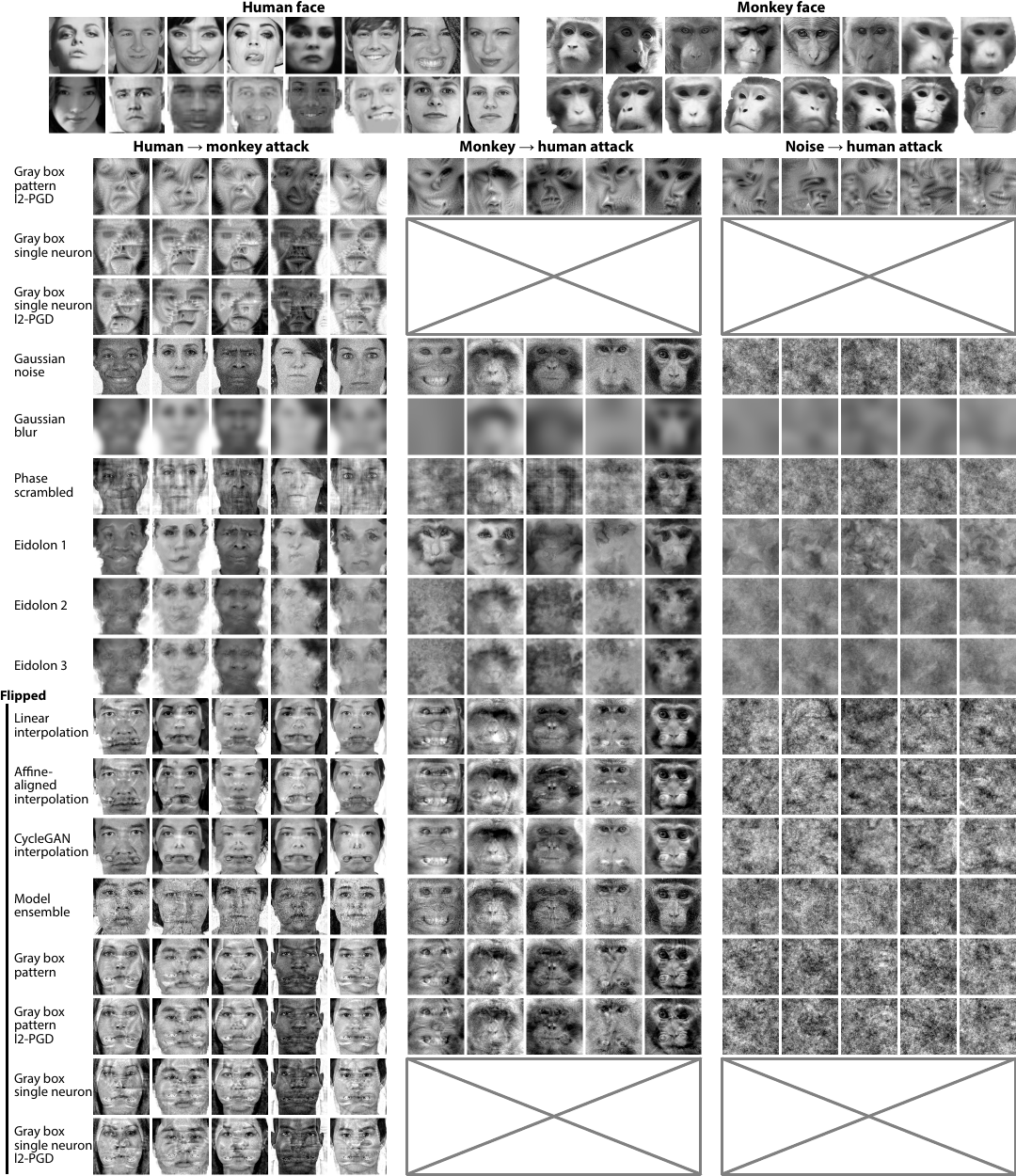}
\caption{Related to \textbf{Figure~\ref{fig:overview}}. Shown are examples of images used in this study: human faces, and level 10 images from additional targeted or control methods not included in \textbf{Figure~\ref{fig:overview}}.}
\label{fig:image_examples}
\end{figure}

\begin{figure}[hpt]
\centering
\hspace*{-0.25in}
\includegraphics{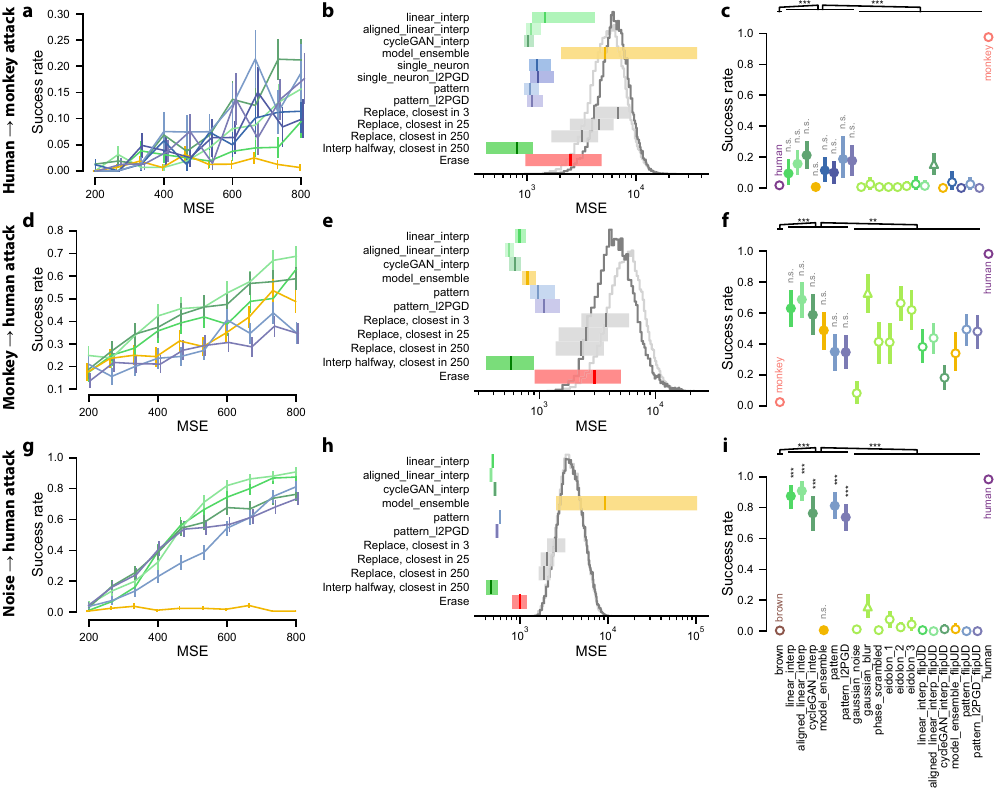}
\caption{Related to \textbf{Figure~\ref{fig:neural}}. Each row indicates a different targeted perceptual change direction. \textbf{a,d,g} show mean success rate and bootstrap 95\%-CI for each method and noise level without logistic regression. In \textbf{b,e,h}, horizontal bars show the MSE level at which success rate for each method should be 50\%. Top bars are based on logistic regression on data from targeted image manipulation methods over levels, and indicate either interpolated (inside tested MSE range of 200--800) or extrapolated (outside tested range) values. Bottom five bars indicate the MSE level associated with image manipulations that were not tested. `Replace' means replacing an original image with the closest target-class example over varying numbers of target-class examples. 'Interp. halfway' means 50\% interpolation towards the closest target-class example. Erase means converting the image to a uniform image with the same mean (which minimizes MSE over uniform images of different images). Vertical tick in each bar shows the sample estimate and extent of the bar shows 95\%-CI based on repeating each analysis over bootstrap samples of images. Two underlying histograms indicate the distribution of pairwise image distances between the target class and clean images from the source class (light gray), or exact same starting images for generating deceptive images from the source class (dark gray). `Replace' calculations were based on the exact starting images. \textbf{c,f,i} shows level 10 success rate of all targeted manipulation methods and control manipulations, as well as accuracy on clean images (with perfect accuracy being 0 and 1 for the source and target category, respectively).}
\label{fig:neural_sup1}
\end{figure}
\clearpage

\begin{figure}[hpt]
\centering
\thisfloatpagestyle{empty}
\includegraphics[scale=0.7]{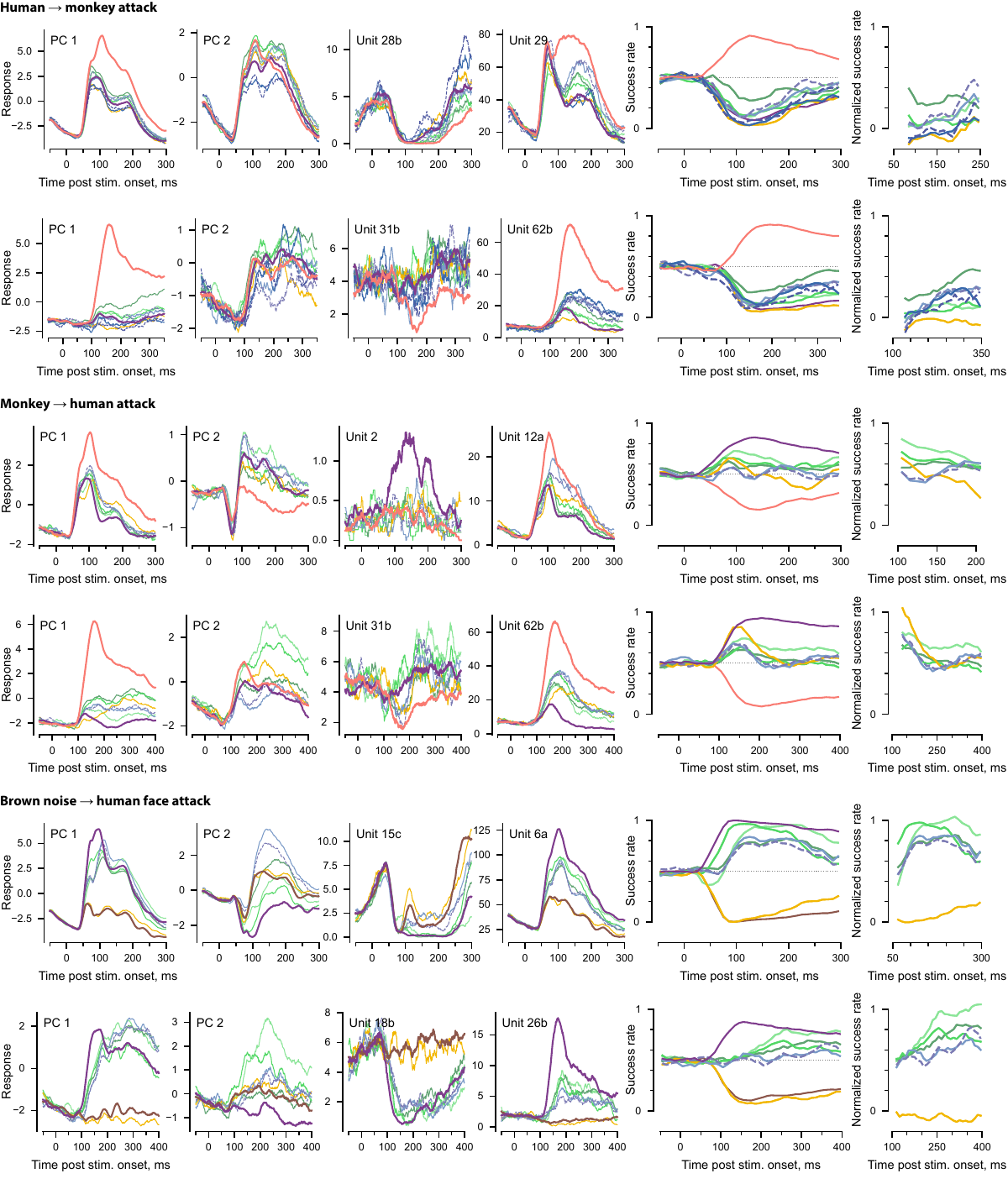}
\caption{Related to \textbf{Figure~\ref{fig:neural}}. Detailed responses over time are shown for clean and targeted manipulated images at level 10. Colors indicate image category as in \textbf{Figure~\ref{fig:neural}}; dashed lines further distinguish $l_2$-PGD methods from non $l_2$-PGD methods, which have similar color. Each row indicates one experimental session. Odd roles indicate one monkey and even rows indicate the second monkey. First four columns show four separate dimensions in neuron responses: 1) first principal component of the population response; 2) second principal component; 3) the most selective unit for one category (quantified by d-prime); 4) the most selective unit for the other category. Y-axis unit is arbitrary for columns 1 and 2, and in units of spikes/s for columns 3 and 4. Column 5 shows success rate quantifying by fitting and testing SVMs on rolling, non-overlapping 10-ms windows of neuronal responses. Column 6 shows the same data as in column 5 with success rates for the clean categories normalized to 0 and 1, showing only the time window during which both clean category accuracies were over 0.75. }
\label{fig:neural_sup2}
\end{figure}
\clearpage


\begin{figure}[hpt]
\centering
\hspace*{-0.25in}
\includegraphics{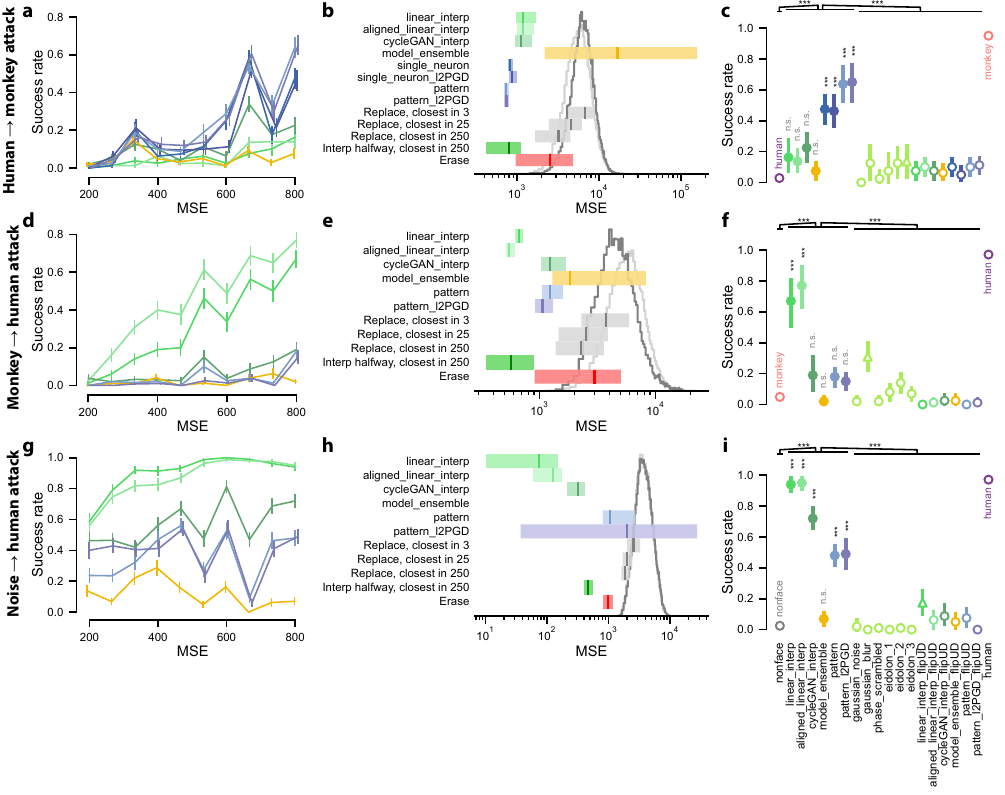}
\caption{Related to \textbf{Figure~\ref{fig:mturk}}. Format of this figure is the same as \textbf{Figure~\ref{fig:neural_sup1}}.}
\label{fig:mturk_sup}
\end{figure}
\clearpage

\begin{figure}[hpt]
\centering
\includegraphics{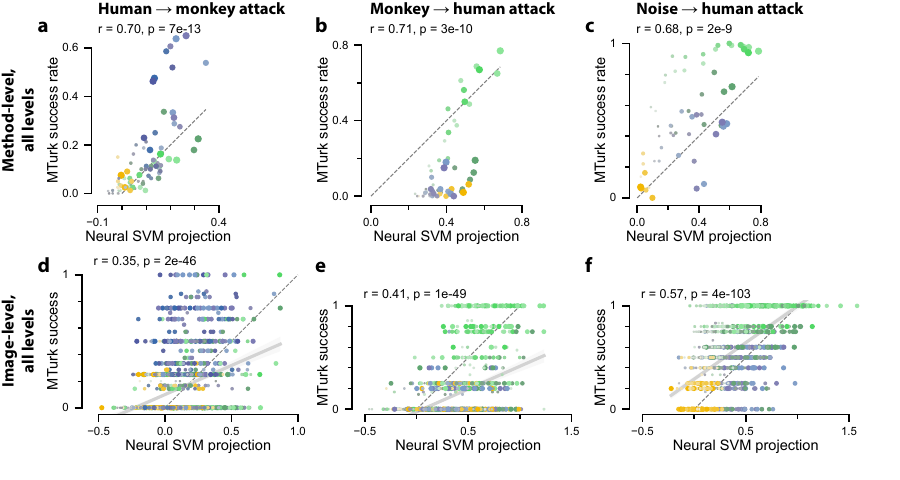}
\caption{Related to \textbf{Figure \ref{fig:compare_hm}} and with similar format. This figure includes images at other noise levels, indicated by dot size.}
\label{fig:compare_hm_sup}
\end{figure}
\clearpage

\begin{figure}[hpt]
\centering
\includegraphics{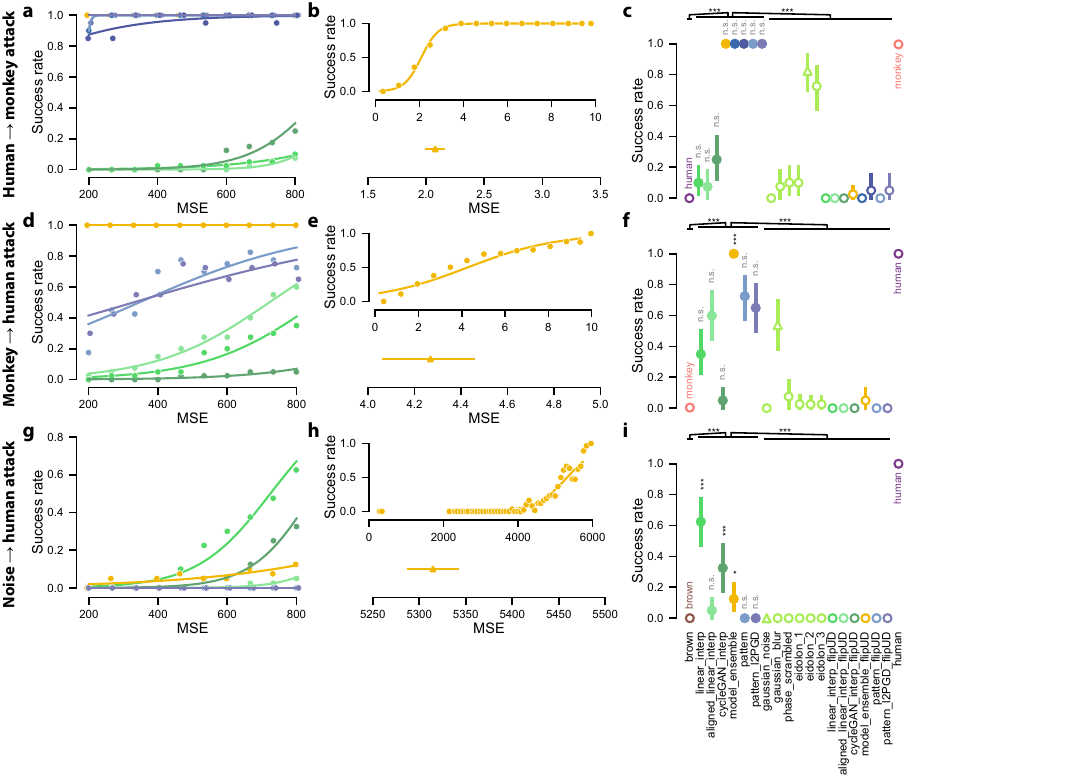}
\caption{Related to \textbf{Figure \ref{fig:model}}. Each row indicates a targeted perceptual change direction. Column 1 shows success rate on an ensemble of CNNs by each image manipulation method at each noise level; format is the same as in \textbf{Figure~\ref{fig:neural}b}. Column 2 top shows the sucess rate of the model ensemble method as a function of MSE for a different range covering low-to-high success rate; bottom shows the estimated threshold for 50\% manipulation success based on logistic regression. Column 3 shows the success rate of all targeted manipulation methods, control image manipulations, and clean images; format is the same as in \textbf{Figure~\ref{fig:neural_sup1}c}.}
\label{fig:model_sup}
\end{figure}
\clearpage

\begin{figure}[hpt]
\centering
\includegraphics[width=1.0\textwidth]{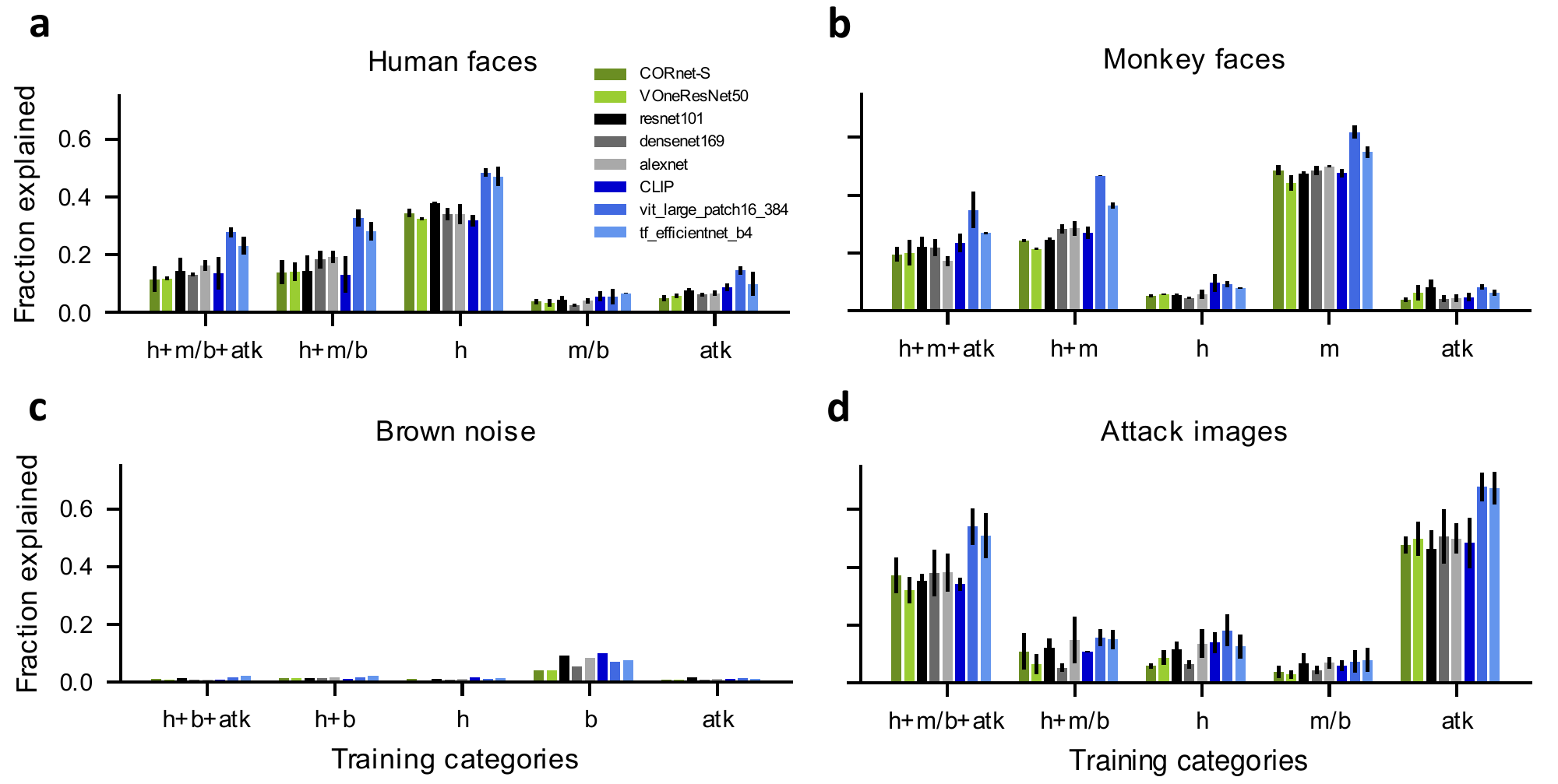}
\caption{Related to \textbf{Figure \ref{fig:model}e--f}. The interpolation and generalization performance in eight ANN models fitted to predict neuronal responses. We tested three `classical' CNN models (ResNet-101, DenseNet-169, and AlexNet), three state-of-the-art ANN models (Visual Transformer, EfficientNet\cite{xie_noisystudent}, and CLIP\cite{radford_CLIP}), and two biologically inspired models (CORnet\cite{KubiliusSchrimpf2019CORnet} and VOneNet\cite{dapello2020simulating}). The bar plots illustrate model performance in different training and testing configurations. Each plot corresponds to testing on one category. Each condition on the x-axis corresponds to one combination of training categories.}
\label{fig:generaliz1}
\end{figure}

\begin{figure}[hpt]
\centering
\includegraphics[width=1.0\textwidth]{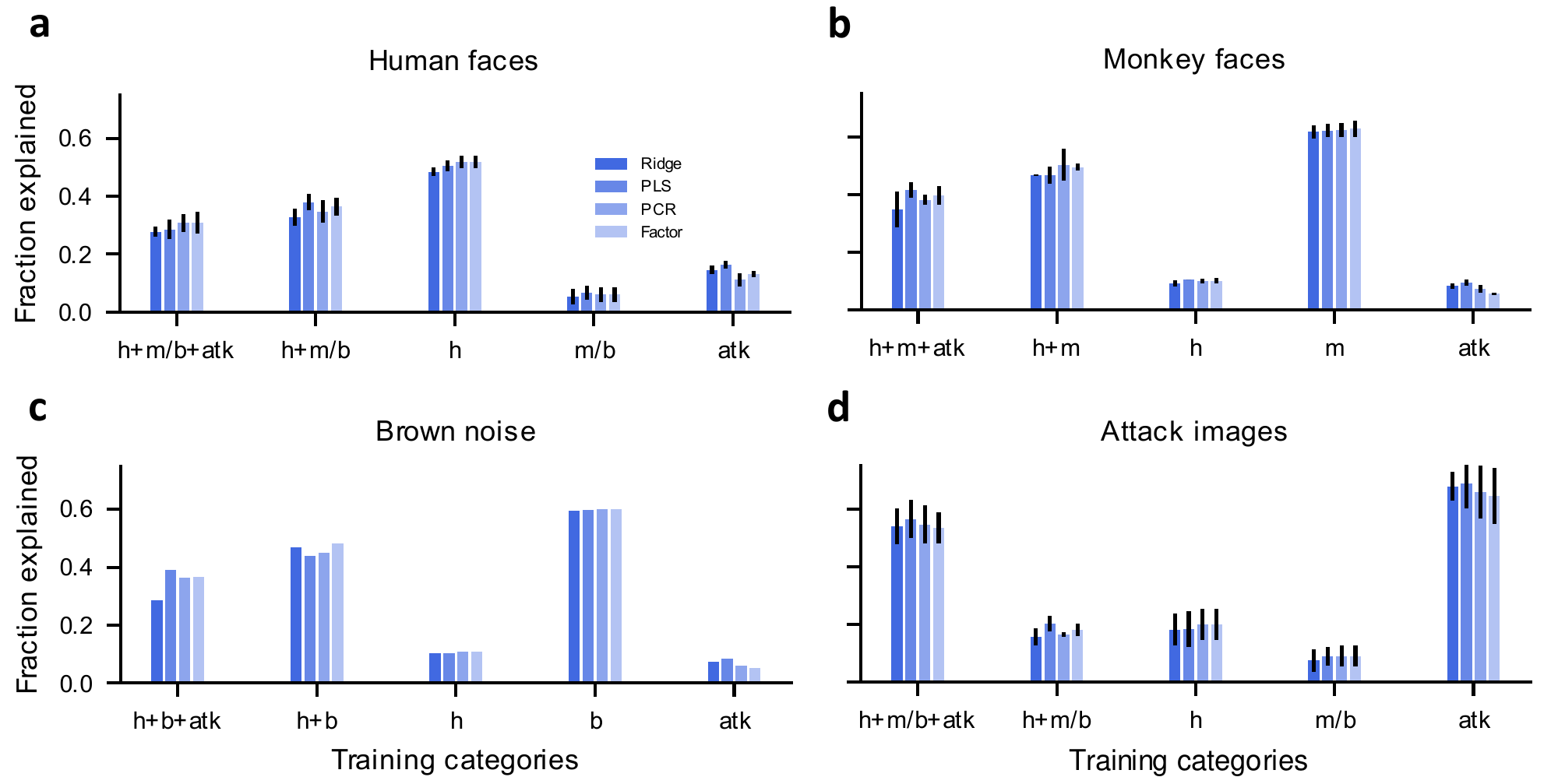}
\caption{Related to \textbf{Figure \ref{fig:model}e--f}. The generalization performance of the Visual Transformer-based predictive model of neuronal responses was similar across four regression methods. We tested ridge regression (alpha = 2 × 10\textsuperscript{5}), partial least squares (PLS, number of components = 27), principal component regression (PCR, number of components = 200), and factor analysis followed by linear regression (Factor, number of components = 250). The bar plots illustrate model performance  in different training and testing configurations. Each plot corresponds to testing on one category. Each condition on the x-axis corresponds to one combination of training categories.}
\label{fig:generaliz2}
\end{figure}
\clearpage


\end{document}